\documentclass[nonacm,sigconf]{acmart} 

\usepackage{amsmath,amsfonts,bm}
\usepackage{graphicx}
\usepackage{url}
\usepackage{textcomp}
\usepackage{verbatim}
\usepackage{array}
\usepackage{enumitem}
\usepackage{wrapfig}
\usepackage{balance}
\usepackage{color,xcolor}
\definecolor{darkblue}{RGB}{0,0,140}  
\usepackage{setspace}
 \usepackage{stfloats}
 \usepackage{etoolbox}

\usepackage{caption}
\usepackage{subfig}
\usepackage{rotating}
\usepackage{tikz}
\usepackage{multirow}
\usepackage{booktabs}
\usepackage{colortbl}
\usepackage{makecell}

\usepackage{algorithm}
\usepackage{listings}

\definecolor{mygreen}{rgb}{0,0.6,0}

\def\BibTeX{{\rm B\kern-.05em{\sc i\kern-.025em b}\kern-.08em
  T\kern-.1667em\lower.7ex\hbox{E}\kern-.125emX}}

\usepackage{hyperref}   
\hypersetup{
    colorlinks=true,    
    linkcolor=blue,     
    citecolor=blue,     
    urlcolor=blue       
}

\newcommand{\scheme}{\textsc{GateBleed}}

\AtBeginDocument{%
  \providecommand\BibTeX{{%
    Bib\TeX}}}

\copyrightyear{2025}
\acmYear{2025}
\setcopyright{acmlicensed}\acmConference[MICRO '25]{58th IEEE/ACM International Symposium on Microarchitecture}{October 18--22, 2025}{Seoul, Republic of Korea}
\acmBooktitle{58th IEEE/ACM International Symposium on Microarchitecture (MICRO '25), October 18--22, 2025, Seoul, Republic of Korea}
\acmDOI{10.1145/3725843.3756097}
\acmISBN{979-8-4007-1573-0/2025/10}

\begin{document}
\pagestyle{plain}


 \title{\scheme: A Timing-Only Membership Inference Attack,  MoE-Routing Inference, and a Stealthy, Generic Magnifier Via  Hardware \textsc{Power Gating} in AI Accelerators} 

\begin{abstract}
As power consumption from AI training and inference continues to increase, AI accelerators are being integrated directly into the CPU. Intel's Advanced Matrix Extensions (AMX) is one such example, debuting in the 4th Generation Intel Xeon Scalable CPU, attaining significant gains in the metrics of performance/watt and decreased memory offloading penalty. This paper discloses a timing side and covert channel, \scheme{}, caused by the aggressive power gating utilized to keep the CPU within operating limits.

This paper shows that the \scheme{} side channel is a threat to AI privacy, as many ML models such as Transformers and CNNs make critical computationally-heavy decisions based on private values like confidence thresholds and routing logits. Timing delays from the selective powering down of AMX components mean that each matrix multiplication is a potential leakage point when executed on the AMX accelerator. This paper identifies over a dozen potential gadgets across popular ML libraries (Hugging Face, PyTorch, TensorFlow, etc.), revealing that they can leak sensitive and private information, including class labels and internal states. \scheme{} poses a risk for local and remote timing inference, even under previous protective measures. 

\scheme{} gadgets can also be used as a \textcolor{black}{a generic} high performance, stealthy 
magnifier for \textcolor{black} {microarchitectural attacks to bypass timer resolution coarsening defenses and create robust and realistic side channels in noisy environments, such as remote attacks on networks with high traffic. This paper shows that when \scheme{} gadget is used as a transmission channel for Spectre, it can leak arbitrary memory addresses of the victim with high performance (0.067 bps), and evade the state-of-the-art microarchitectural attack detectors for the first time.}  



\textcolor{black}{This paper implements an end-to-end membership inference attack with 81\% accuracy on a Transformer model optimized with Intel AMX and 99\% accuracy on an early-exit CNN classifier. \scheme{} achieves 0.89 precision while leaking expert choice in a Transformer mixture-of-experts (MoE) with 100\% accuracy. These attacks do not rely on confidence scores or model outputs, but only on the execution time of attacker-controlled AMX instructions on the shared hardware accelerator with power gating.}
%
To the authors' knowledge, this is the first side-channel attack on AI privacy that exploits hardware \textcolor{black}{accelerator power}  optimizations. \textcolor{black}{The paper also suggests effective mitigations and measures their trade-off between \textit{power consumption} and \textit{performance}. }

\end{abstract}
\author{Joshua Kalyanapu}
\email{jkalyan@ncsu.edu}
\affiliation{%
  \institution{NC State University}
  \city{Raleigh}
  \state{NC}
 \country{USA}
}

\author{Farshad Dizani}
\authornote{Both authors contributed equally}
\email{fdizani@ncsu.edu}
\affiliation{%
  \institution{NC State University}
  \city{Raleigh}
  \state{NC}
  \country{USA}
}

\author{Darsh Asher}
\authornotemark[1]
\email{dkasher@ncsu.edu}
\affiliation{%
  \institution{NC State University}
  \city{Raleigh}
  \state{NC}
 \country{USA}
}

\author{Azam Ghanbari}
\email{aghanba2@ncsu.edu}
\affiliation{%
  \institution{NC State University}
   \city{Raleigh}
   \state{NC}
  \country{USA}
}

\author{Rosario Cammarota}
\email{rosario.cammarota@intel.com}
\affiliation{%
  \institution{Intel Labs}
   \city{San Diego}
   \state{CA}
  \country{USA}
}

\author{Aydin Aysu}
\email{aaysu@ncsu.edu}
\affiliation{%
  \institution{NC State University}
   \city{Raleigh}
   \state{NC}
  \country{USA}
}

\author{Samira Mirbagher Ajorpaz}
\email{smirbag@ncsu.edu}
\affiliation{%
  \institution{NC State University}
   \city{Raleigh}
   \state{NC}
  \country{USA}
}


\maketitle


\section{Introduction}
\label{sec:introduction}


AI applications are power intensive~\cite{openai2018,garcia2019estimation}. Advanced power optimizations such as power gating are being deployed in AI accelerators to reduce the power consumption overhead and ease this burden~\cite{kandiah2021accelwattch,kumar2014efficient,wang2011power,lungu2009dynamic,arora2015understanding,kosonocky2011practical}, however, their privacy and security risks on AI applications and remote settings are poorly understood.

As we enter a new era of computing, each leap in processor technology not only increases performance and efficiency, but alters the landscape of cybersecurity threats~\cite{pandorasbox}. 
Microarchitectural optimizations can introduce side channels that can expose sensitive data to adversaries.
Spectre \cite{kocher2020spectre} and Meltdown \cite{meltdown} show how speculative and out-of-order execution can be exploited as side channels exposing private data through subtle timing differences. Subsequent research uncovered an avalanche of vulnerabilities exploiting 
various microarchitectural components, such as 
the cache ~\cite{flush+reload, gruss2016flush+, osvik2006cache}, internal CPU buffers ~\cite{kocher2020spectre,Phantom2023,spectre_here_to_stay2019, SpectreV4,kiriansky2018speculative,koruyeh2018spectre,maisuradze2018ret2spec,meltdown, canella2018systematic,van2018foreshadow,lvi, moghimi2020medusa,ridl,
schwarz2019zombieload, canella2019fallout, moghimi2023downfall,CrossTalk2021,weber2021osiris}, prefetchers ~\cite{gruss2016prefetch,lipp2022amd,augury,Chen2024GoFetch}, pointer authentication~\cite{PACMAN}, TLB ~\cite{Tatar2022tlbdr,gras_translation_2018,koschel2020tagbleed}, execution ports~\cite{bhattacharyya2019smotherspectre,port-contention}, scheduler queues~\cite{SQUIP2023}, micro-op cache~\cite{uop2021Ashish}, RAPL~\cite{Lipp2021Platypus}, and DVFS~\cite{wang2022hertzbleed,wang2023dvfs,liu2022frequency},
showing that hardware-based timing side channels undermine the very foundations of higher-level isolation and trust—rendering traditional defenses such as encryption, sandboxing, memory isolation, and privilege separation insufficient in the face of hardware-level information leakage.
 MLaaS interfaces and multi-tenant cloud services are increasing in deployment, but little attention has been paid to the impact of optimizing hardware accelerators on the security and privacy of AI applications, as well as user privacy. 
 Previous works focus on attacks that extract 
 architectures and hyperparameters 
~\cite{TramerZJRR16,Orekondy18Knock,jagielski2020highaccuracyhighfidelity,Chandrasekaran2020Active,wang2018hyperparameters,Yan2020CacheTelepathy},
alongside inference attacks that use output confidences or logits and 
training a proxy model that behaves similarly to the under-attack model using the under-attack model's output 
to deduce model properties
such as training-set membership (MIAs) \cite{Shokri2017SP,Salem2018MLLeaks,hu2022membership,li2021membership,he2025towards,Yeom2018CSF,choquette2021label,hilprecht2019monte,gupta2021membership,duddu2020quantifying,liu2021encodermi,liu2019performing}, attributes of user inputs \cite{zhao2021feasibility,gong2018attribute,mehnaz2022your,JayaramanAtt}, and 
decisions in adaptive models such as early exits and MoE routing \cite{teerapittayanon2016branchynet,msdnet,Brennan2020SP,Li2022CCS,ding2025moecho}.
%
Restricting model outputs to labels rather than confidence scores~\cite{Shokri2017SP} is largely effective against many membership inference attacks.
Differentially private training~\cite{abadi2016deep,Jayaraman2019DPML} prevents a single training sample from significantly affecting the output. Finally, execution padding with dummy instructions in adaptive DNNs hides dynamic routing and early‑exit timing\cite{Brennan2020SP,Li2022CCS,Akinsanya2024TimingCI}. However, no work has studied whether these leaks can originate below the API \textemdash in hardware optimization itself, where higher-level isolation, such as output masking and DP training, is utterly insufficient to secure the data.

\definecolor{myblue}{HTML}{007799}
\definecolor{myorange}{HTML}{FF8888}
\begin{figure} [!htbp]
    \centering
    \includegraphics[width=\linewidth]{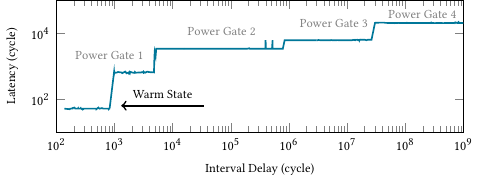}
        
         
    \caption{  Performance States of
TMUL due to Power Gating Leaks Private AI Parameters or Arbitrary Data and the Secure Recommended Intel AMX Operational State (Warm)}
    \label{fig:Performance_Stages}
\end{figure}

Conventional timing side-channels that exploit hardware optimizations to target AI privacy leak data 
stored in memory or in registers in the form of bits, not properties of the data. 
For example, Cache Telepathy~\cite{Yan2020CacheTelepathy} uses cache-based timing to recover model architectures and hyperparameters present in memory. Hertzbleed~\cite{wang2022hertzbleed} or Platypus~\cite{Lipp2021Platypus} use operand-dependent CPU frequency throttling to leak register contents. 
On the other hand, \emph{AI privacy attacks} obtain 
information that is \emph{not} stored as bytes, e.g.,\ 
training-set membership or expert routing choice. The leaked information exists only as latent properties of the learned model, and in some cases, the execution leaves observable footprints on hardware. This paper investigates the extraction of properties of the data used to train the AI model via timing discrepancies due to the \emph{hardware optimization} of on-core accelerator power gating.

With the sharp increase in power consumption required by modern AI~\cite{openai2018,garcia2019estimation}, hardware designers have been looking to further increase the energy efficiency of AI inference and training. 
Intel recently introduced Advanced Matrix Extensions (AMX), an on-core AI accelerator available in 
Scalable Xeon CPUs
~\cite{nassif2022sapphire,munch20242,varada20252}. AMX achieves up to 10$\times$ inference and training performance and 3$\times$ performance per watt improvement over the previous generation. AMX delivers high throughput for the multiplication of int8 and bfloat16 matrices, achieving more than 30 TFLOPS by executing up to 1024 MACs per cycle, culminating in up to a 75\% lower Total Cost of Ownership (TCO) than the previous generation of Scalable Xeon CPUs~\cite{sprproductbrief}.
However, no prior works have explored the security and privacy implications of Intel AMX. 

We performed a complete characterization of AMX performance and power and found novel timing channels in the form of a \textit{reuse-distance-dependent} timing difference of up to 20,000 cycles with a single AMX instruction (see Figure~\ref{fig:Performance_Stages}). This difference is even visible when AMX is run inside an Intel SGX secure enclave~\cite{SGX}, allowing observation of AMX utilization even within a trusted execution environment. 
We found the root cause of the high timing discrepancy to be an undisclosed power gating optimization, a ubiquitous power-saving measure that entails the powering off of unused chip components to save power~\cite{lungu2009dynamic,arora2015understanding,kosonocky2011practical}. While power-gating does reduce power consumption, power-gating also incurs an extra measurable latency when waking up a powered-off component, hence enabling a novel reuse-distance-dependent timing difference.  We thus introduce a side-channel attack, \scheme{}, that \textit{bleeds} information via the wakeup latency induced by power \textit{gating}.



By monitoring accelerator timing, we can determine training‑set membership with high accuracy. 
%
The ability to determine whether specific data were used to train a deployed model has implications for user privacy, security, and corporate liability. Leaking membership in an ML model reveals whether a specific input was part of the training set, directly threatening data privacy by violating confidentiality guarantees of MLaaS systems. Attackers who know a model’s training corpus can better craft adversarial or targeted attacks \cite{TransferabilityIM2016,aghakhani2021bullseye,CarliniPoisoning}. Recent lawsuits involving Stability AI~\cite{GettyImages_v_StabilityAI_2025_EWHC38_Ch}, Google~\cite{ICO_Guidance_on_AI_Data_Protection_2023,EDPB_Opinion282024}, and Meta~\cite{veale2018algorithms} highlight that companies may face legal exposure if one can 
demonstrate that proprietary or unauthorized data were used during training~\cite{ICO_Guidance_on_AI_Data_Protection_2023,EDPB_Opinion282024,GettyImages_v_StabilityAI_2025_EWHC38_Ch,Reuters2025AppleBooksAILawsuit,ThomsonReuters_v_RossIntelligence_2025_DDELD,WP2025AnthropicFairUse,veale2018algorithms}.

Membership inference attacks (MIAs) exploit the tendency of models to be more confident in training examples than in unseen examples. In the original black‑box MIA~\cite{Shokri2017SP}, the adversary trains shadow models that mimic the target and also trains a meta‑classifier using the shadow model outputs (labels and confidences) for known members and non‑members; the meta‑classifier is then applied to the target model’s outputs. Subsequent work extended MIAs beyond classifiers and to more restrictive APIs, including attacks on non-classifier models~\cite{hayes2017logan,hilprecht2019monte,liu2019performing,gupta2021membership,duddu2020quantifying,liu2021encodermi}, models that use only top‑K confidences~\cite{Salem2018MLLeaks}, or even label‑only outputs~\cite{he2025towards,Yeom2018CSF,li2021membership,choquette2021label,rahimian2020sampling}. In practice, the strongest MIAs generally rely on access to logits, which are often hidden in production systems. 
Such leakages threaten data privacy and may raise compliance concerns~\cite{murakonda2020ml}. However, many MIAs degrade over noisy channels (e.g., real‑world networks) and become less effective when only hard labels are exposed. To our knowledge, \scheme{} is the first timing side-channel attack utilizing a hardware accelerator power optimization to target user data privacy without relying on the confidence scores or even an output label.

Leveraging the insight that ML models tend to exhibit higher confidence on training-set members than non-members~\cite{Akinsanya2024TimingCI}, we utilize \scheme{} to detect AMX usage of an adaptive neural network running on the same core as an attacker-controlled program; the AMX usage patterns allow the attacker to infer whether or not the adaptive neural network was trained on a particular input. Inputs that were in the 
training set are more likely to produce high-confidence predictions~\cite{Shokri2017SP}, triggering different behavior in adaptive neural networks~\cite{Akinsanya2024TimingCI}. Adaptive neural networks have been deployed in 
 products by Google~\cite{leviathan2023fast}, Meta~\cite{khandelwal2021jointly}, Amazon~\cite{Trimbach2025}, and Alibaba~\cite{liu2017cascade}. 
We show that by detecting early exits or smaller expert execution, the attacker infers membership without accessing logits, output probabilities, or model internals, compromising user data privacy through power gating.

\scheme{} can also infer private decisions such as expert routing in mixture‑of‑experts transformers by correlating accelerator timing with model behavior. In particular, 
\textit{Mixtral (HF)}~\cite{huggingface_transformers}, \textit{TensorFlow MoE}~\cite{tensorflow}, \textit{DeepSpeed MoE}~\cite{deepspeed_moe}, \textit{ONNX RunTime MoE}~\cite{onnxruntime_moe}, and \textit{Mixtral (llama.cpp)}~\cite{llama_cpp_ggerganov} exhibit AMX-backed conditional expert execution that is exploited in our ~\scheme{} attack.
For example, recent works have proposed heterogeneous 
experts~\cite{wang2024hmoe} or a different number of experts activated for each token in the Transformer~\cite{huang2024harder}, leading to an observable AMX timing state. 
\scheme{} harnesses this difference in computation to leak the \textit{expert routing index} through the power gating footprints. 
This work identifies more than a dozen exploitable timing gadgets across popular machine learning stacks (including PyTorch, TensorFlow and HuggingFace implementations), showing the footprint of accelerator power gating in modern AI pipelines and agents. 
\scheme{} enables a broader family of accelerator-driven privacy leaks. Any residency- or power-managed compute path—on-core AI engines (e.g., AMX/NPUs), GPU tensor cores, or dedicated inference IP—can imprint input- or model-dependent timing. The resulting signals need not expose stored bytes; they can encode decisions about membership, routing, tool selection in Agentic AI, or early exits across processes, VMs, and TEEs, and can be magnified for remote settings.

Because ~\scheme{} exploits hardware design, the vulnerability cannot be fixed with a simple software update. Effective mitigation likely requires hardware changes that take years to reach deployed systems and have high power overhead. 
Because ~\scheme{} doesn’t rely on model outputs such as confidence scores, \scheme{} evades existing defenses built for inference attacks.
For example, proposed software-level defenses such as 
confidence score masking~\cite{Shokri2017SP} and differential privacy 
~\cite{dwork2006calibrating} do not work to mitigate ~\scheme{}. 

Furthermore, this paper explores using ~\scheme{} as the covert channel for a remote Spectre-v1~\cite{kocher2020spectre} and as a microarchitectural magnifier to measure a microarchitectural state in an environment in which only a coarse timer is available. The remote Spectre-v1 attack achieves never-before-seen stealthiness and leakage rate on a noisy, production network while bypassing state-of-the-art microarchitectural attack detectors like 
Evax~\cite{EVAX2022Micro}, 
PerSpectron~\cite{PerSpectron}, and 
RHMD~\cite{RHMD2017}. Utilizing ~\scheme{} as a magnifier allowed distinguishing between an L1 cache hit or an L3 cache miss using an exceptionally coarse timer. 
The ability to use ~\scheme{} as a general-purpose covert channel and as a microarchitectural magnifier highlights additional threats posed by power-gating optimizations. 

We revisit the state-of-the-art microarchitectural attack detectors and show that they achieve near-100\%
detection of recent magnifiers~\cite{HackyRacers2023ASPLOS, Microscope} with high accuracy due
to the magnifiers' visible execution patterns, reliance on hardware optimizations
(e.g., branch predictor, cache, prefetcher), and high repetition rates.
However, microarchitectural attack detectors fail to detect \scheme{} attack variants in
time, making \scheme{} the only tested microarchitectural side channel attack that evades microarchitectural attack detectors, because detectors (i) do not account for the misuse of
hardware \emph{power} optimizations such as power gating exploited in
this attack, (ii) cannot resist the low repetition rates required for
\scheme{} to succeed, (iii) do not capture the passive reset phase of
\scheme{}, which avoids anomalous microarchitectural instructions, and
(iv) miss accelerator-specific features and counters from their design.



We propose mitigating \scheme{} with a microcode update that maintains AMX in a single stage or using compiler-inserted AMX dummy operations to maintain the AMX power stage. This work shows that these mitigations increase power consumption overhead by up to 12\%, but secures 
against \scheme{}.

\vspace{1em}
\noindent\textbf{Contributions:}  This paper makes the following contributions: \begin{itemize}
    \item We uncover and characterize the undocumented power gating in AMX, revealing five distinct latency states that vary based on time of last usage. 
    \item We present \scheme, the first microarchitectural \emph{privacy} side channel that exploits on-core accelerator power gating to infer training-set membership, MoE routing, and early exits—without logits/confidences or access to stored artifacts.
    We exploit existing benign ML libraries,  
   leaking sensitive private ML secrets and inference paths 
   across OS and VM. 
 \textcolor{black}{ Our end-to-end attack on MoE achieves 100\% success rate with 0\% FPR. 
 Similarly, the early-exit CNN attack
 achieves 99.72\% accuracy with just 0.54\% false positives.
 Our membership inference attack on early-exit Transformers, leveraging the same timing mechanism of~\scheme{}, 
 yields 81\% accuracy (78\% TPR, 84\% TNR).}
    \item We present a variant of \scheme{} as a generic stealthy magnifier in intensifying a 200-cycle timing difference to a 10,000-cycle difference. We evaluated \scheme{} as a covert channel that achieves a 70,000$\times$ higher leakage rate than NetSpectre on our production network, where the prior covert channels fail.  We show that \scheme{} evades malware detectors trained on microarchitectural attack patterns due to minimal instruction footprint and absence of cache or TLB flushing. \scheme{} also defeats timer coarsening defenses by operating with timing margins of up to 20,000 cycles. 
    
    \item We discuss 
     mitigations to close this attack vector and measure the mitigations' power and performance overhead, concluding that the most practical solution, powering off AMX on a context switch, results in a power consumption overhead between 2\% and 12\% depending on the context switch rate. 
\end{itemize}

\noindent\textbf{Responsible Disclosure:} We disclosed this issue to Intel from May 2023 to May 2024. Intel confirmed our findings, and Lenovo released a firmware mitigation. In June 2024, Lenovo released a UEFI update, Version 3.20, Build ID ESE126H [Critical], which mitigates the most critical \scheme{} variants, particularly those exploiting AMX power gates 3, 4, or 5. 
\noindent We have open-sourced the \scheme{} proofs-of-concept (PoCs)\footnote{Code available at
\url{https://zenodo.org/records/17019733}. All results in this paper are
reproducible within 40 minutes following the instructions at
\url{https://github.com/jkalya/gatebleed}, and have been verified through the MICRO artifact evaluation.}.




\vspace{1em}
\noindent \textbf{Paper Organization:} 
Section ~\ref{sec:background} provides background on timing channels, timing channel defenses, adaptive neural networks, and AI model vulnerabilities. Section~\ref{sec:threat} presents the threat model and attack requirements. Section~\ref {sec:overview} presents a detailed overview of Intel AMX architecture and the \scheme{} vulnerability. Section ~\ref{sec:attack} introduces the \scheme{} attack, demonstrating its use in side channels, covert channels, and magnification gadgets across real ML workloads. Section~\ref{sec:results} provides the practical results. Section ~\ref{sec:defense} proposes effective mitigations and evaluates their overhead. Finally, Section ~\ref{sec:conclusion} concludes the paper.

\section{Background}
\label{sec:background}

\noindent\textbf{Timing Microarchitectural Attacks.}
Covert channels exploit shared microarchitectural states to transmit information using timing differences. Classical cache-based attacks, such as \emph{Prime+Probe}~\cite{osvik2006cache}, \emph{Flush+Flush}~\cite{gruss2016flush+}, and \emph{Flush+Reload}~\cite{flush+reload}, rely on eviction or access timing to leak fine-grained memory behavior. TLB-based attacks such as \emph{TLBleed}~\cite{gras_translation_2018} and \emph{Binoculars}~\cite{zhao2022binoculars} extend cache-based covert channels to memory translation structures, using page-level state or page-walker contention. Branch predictor attacks~\cite{Evtyushkin2018BranchScope} infer secret control flow through the predictor state, typically requiring simultaneous multithreading (SMT).
Transient attacks such as Spectre~\cite{kocher2020spectre} and Meltdown~\cite{meltdown}, and MDS-type attacks~\cite{kocher2020spectre,meltdown, ridl,schwarz2019zombieload} exploit microarchitectural optimizations such as out-of-order execution or speculation to access unauthorized data transiently; speculative execution attacks rely on benign code that has unintended effects on the hardware. Values leaked in a speculative execution attack are exfiltrated via a covert channel. Hardware mitigations like speculative barriers and buffer flushing limit, but do not eliminate such attacks. Remote variants like Netspectre~\cite{netspectre} demonstrated feasibility even without attacker code on the victim machine, albeit at extremely low bandwidth, which fails on a real production network because of the network noise masking timing differences. 
%

\noindent\textbf{Power and Frequency Optimization Based Attacks.}
\emph{Platypus}~\cite{Lipp2021Platypus} exploits the RAPL interface to leak secrets from SGX enclaves via data-dependent power consumption. Hertzbleed~\cite{wang2022hertzbleed} shows how data-dependent power consumption can leak information via timing due to CPU frequency throttling. 
Wang et al.~\cite{wang2023dvfs} extend frequency throttling via data-dependent power consumption to on-board graphics. 
Rauchscher et al. ~\cite{rauscher2024idleleak} exploit CPU sleep states, or C-states, to introduce timing differences. Thor~\cite{dizani2025thor} exploits an operand-dependent timing difference observed in Intel AMX's lowest power state to leak operand sparsity in matrix multiplications. These channels are powerful but vulnerable to noise and can be suppressed through frequency locking, access restrictions, added noise to power measurements~\cite{Lipp2021Platypus}, and reduced measurement rate. 


\noindent\textbf{Microarchitectural Magnifiers.}
Microarchitectural magnifiers are techniques to magnify a microarchitectural effect that is hard to measure with conventional means. Hacky Racers~\cite{HackyRacers2023ASPLOS} constructs ILP-based magnification gadgets to induce a potentially unbounded timing difference, defeating timer coarsening defenses such as the standardized 5 $\mu$s timer resolution~\cite{W3C2024_HighResolutionTime3} implemented in Chrome, Firefox, Edge, and Safari and enabling microarchitectural side-channel attacks in the browser which exploit differences on the order of nanoseconds. Bypassing the same timer coarsening, Spring~\cite{wikner2022spring} encodes a secret into multiple cache lines instead of a single one.
Microscope~\cite{Microscope} is a modification to the OS page fault handler attacking trusted execution environments (TEEs) like Intel SGX. Attacks on TEEs assume a privileged attacker. By modifying the page fault handler to constantly retry, small sections of victim code repeat execution, making side-channels that rely on extremely subtle effects like execution port contention~\cite{port-contention} easily visible with only one architectural run of the victim code.
However, existing magnifiers require attacker-controlled code or complex code that is unlikely to exist in a benign codebase; furthermore, our experiments have shown that the existing magnifiers are highly detectable by the state-of-the-art microarchitectural attack detectors~\cite{EVAX2022Micro,PerSpectron}.

\noindent\textbf{Attacks on AI Security and Privacy}. \textit{Data security attacks} target the model artifacts themselves (e.g., parameters, hyperparameters, or proprietary architectures) resident on the victim machine (often in process memory). Adversaries exploit physical and microarchitectural side channels such as cache access patterns~\cite{Yan2020CacheTelepathy}, timing variations~\cite{Duddu2018Timing}, electromagnetic emanations\cite{Batina2019USENIX}, and power consumption~\cite{wei2018know,dubey2020maskednet} to recover sensitive model information or keys, or to guide model extraction. Other avenues include training high‑fidelity surrogates to replicate a target model’s decision boundary~\cite{jagielski2020highaccuracyhighfidelity} and cryptanalytic/model‑extraction techniques that recover parameters or architectures from query accesses or compressed representations\cite{carlini20-extract,carlini2024polynomial}.

\textit{User privacy attacks}, by contrast, exploit input–output behavior to infer sensitive information about the data used to train or query the model—items not directly present in process memory. This includes membership inference\cite{Shokri2017SP,li2021membership,gupta2021membership,liu2021encodermi,liu2019performing,hilprecht2019monte,choquette2021label,he2025towards} and input data attribute inference \cite{zhao2021feasibility,gong2018attribute,mehnaz2022your}. Adaptive or early‑exit DNNs introduce additional leakage channels because inputs may traverse different subnetworks~\cite{teerapittayanon2016branchynet, dubost2019hydranet, msdnet}; routing choices and early exits can be inferred remotely via end‑to‑end latency measurements even when model parameters are fixed~\cite{Brennan2020SP,Li2022CCS}. Timing signals can also leak membership in adaptive networks: overconfident routing can make training members exit earlier than non‑members, yielding faster inference on members~\cite{Akinsanya2024TimingCI}.  These timing‑based attacks typically require precise synchronization and stable network conditions; in contrast, \scheme{} enables an attacker to infer private information by measuring only its own program’s latency on shared hardware, removing the need for end‑to‑end timing of the victim model.

\section{Threat Model}\label{sec:threat}
Here we enumerate the threat models we assume for the \scheme{} attack \textemdash the attack targets and attacker capabilities. Table~\ref{tab:gadgets} shows \scheme{} gadgets present in real-world codebases along with the threat model we assume for leaking with that gadget. 

The \scheme{} \textit{attack target} is any internal model parameter or decision whose unintended disclosure through timing variations can compromise user privacy, model confidentiality, or operational integrity. Specifically, these secret parameters include: (1) intermediate {confidence scores or prediction entropy values};
this leakage enables precise membership inference attacks by distinguishing training-set members with typically lower intermediate entropy from non-members~\cite{Shokri2017SP};  
(2) internal {routing decisions} such as expert selection in Mixture-of-Experts (MoE) models, where the activated expert is determined by private input-dependent logits; and 
(3) operational or contextual flags, such as {session reuse indicators} (key-value cache usage) or {quantization configurations}, revealing sensitive model runtime states or deployment settings. 

\begin{table*}[t]
\centering
\renewcommand{\arraystretch}{1.2}
\footnotesize
\setlength{\tabcolsep}{4pt}
\begin{tabular}{|l|l|p{9.2cm}|}
\hline
\textbf{Library / Model} & \textbf{Leaked Parameter}/ \textbf{Variable} & \textbf{Leakage Path and Threat Model (Query / Timing)} \\
\hline
\multicolumn{3}{|c|}{\textbf{I. Input-Dependent Routing Gadgets}} \\
\hline
Mixtral (HF)~\cite{huggingface_transformers} & Expert routing index/Routing logits & AMX matmuls execute only for selected experts; Query+Time reveals routing decisions. \\ 
AdaptiveLogSoftmax~\cite{pytorch} & Cluster membership/Target label & Cluster-based branches vary in compute time; Query+Time reveals true class. \\
TensorFlow MoE~\cite{tensorflow} & Router activation/Routing threshold & AMX triggered only for active experts; timing reveals routing threshold decisions. \\
DeepSpeed MoE~\cite{deepspeed_moe} & Sparse expert maskMoE gating pattern & Expert selection alters AMX load; Query+Time timing reveals active paths. \\
ONNX Runtime MoE~\cite{onnxruntime_moe} & Active expert path/Expert selection & Expert-specific AMX ops in If-nodes; timing-only observer can infer selected branch. \\
Mixtral (llama.cpp)~\cite{llama_cpp_ggerganov} & Gate threshold/Router logits & First expert triggering AMX leaks routing; fine-grained timing from query reveals selection. \\
RGATConv (PyG)~\cite{pytorch_geometric} & Edge-type dependency/Edge attribute & Conditional projection varies by edge type; timing-only observer reveals structural edges. \\
LangChain~\cite{langchain} & Tool dispatch category/Classification logits & Tool path alters execution time; Query+Time reveals decision path. \\
{ggml~\cite{githubGitHubGgmlorgggml}} & {Batch size} & {Executes AMX only if batch size is not 1. AMX usage leaks if batch size is 1} \\
\hline
\multicolumn{3}{|c|}{\textbf{II. Confidence and Early-Exit Gadgets}} \\
\hline
BranchyNet~\cite{teerapittayanon2016branchynet} & Exit stage decision/Confidence score & Exit stage changes AMX pattern; Query+Time/AMX usage reveals model confidence. \\
MSDNet~\cite{msdnet} & Early-exit threshold/Prediction entropy & Prediction entropy modulates early-exit logic. Query+Time reveals entropy. \\ 
SkipNet / BlockDrop~\cite{skipnet} & Layer skipping pattern/Routing mask &  Conditional skips affect AMX reuse. Latency pattern encodes layer execution mask. \\
HF Agent~\cite{huggingface_transformers} & Action type/Action logits & Decoder used only for tool tokens; timing from query reveals action category. \\
AutoGen~\cite{autogen_microsoft} & Loop interval/Planner state & Internal planner invokes AMX conditionally. Timing leaks planning state. \\ 
\hline
\multicolumn{3}{|c|}{\textbf{III. Session, Configuration, and Static Context Gadgets}} \\
\hline
LLaMA KV Cache~\cite{huggingface_transformers} & KV reuse vs. recompute/Context reuse & Reused prompt avoids recomputation; query+timing reveals reuse and leaks prompt history. \\
ONNX Runtime KV Cache~\cite{onnxrunTime_kvcache} & Session persistence/Session reuse & Warm sessions reduce AMX setup time; timing-only observer detects session reuse. \\
llama.cpp Quant Dispatch~\cite{llama_cpp_ggerganov} & Model format/Quantization type & Model quantization (int8 vs fp32) toggles between AMX/AVX. Timing reveals format. \\ 
GoogLeNet~\cite{pytorch}  & Training mode/Training flag & Auxiliary classifier invoked only in training. Latency reveals execution mode (train vs infer). \\ 
Generic CNN~\cite{Kerasteamkeras} & Layer type / Architecture toggle & 
Architecture flagged (e.g., conv vs MLP) alter AMX invocation. Latency exposes layer type. \\ 
OpenAI Function API~\cite{openai_function_calling} & Completion state/Finish signal & Completion path triggered AMX only for function tool. Latency leaks endpoint behavior. \\ 
\hline
\end{tabular}

\caption{Categorized \textsc{GateBleed} gadgets and threat models. Each row describes a parameter that influences AMX usage and may leak via timing or query-time side channels. The final column summarizes the leakage path and attacker model.}
\label{tab:gadgets}
\end{table*}

Attacker capabilities vary depending on the target gadget. The "Threat Model" column in Table~\ref{tab:gadgets} highlights the range of attacker capabilities required to exploit each gadget. Namely, we investigate three threat models - (1) \textit{Query+Time} where the attacker remotely queries a MLaaS model and times the response time to leak details about the model, (2) \textit{Timing} where the attacker remotely sniffs packets on the network to determine the response time of a user's request to the MLaaS model to leak details about the user's input, and (3) \textit{AMX Usage} where the local attacker colocated on the same core as the MLaaS program can both induce the model to run and time its own AMX operations to leak. 
   
The most constrained \textit{remote} attackers (Query+Time, Timing threat models) can leak several targets. 
   For example, an adversary over the network can send carefully chosen queries to an MLaaS API and measure response times to infer facts about the model. This minimal attacker model only needs to observe overall request latency, and \scheme{}'s signal remains detectable despite network noise because of the significant timing difference. The attacker does not know model weights, architectural parameters, or outputs and is unprivileged, but using response latencies can build a correlation of inputs to response times. 


   Gadgets become even more pronounced if the attacker is on the same host as the MLaaS model due to the ability to obtain a more fine-grained view of AMX usage than what end-to-end timing reveals (AMX Usage threat model). 
   \textcolor{black}{
Assuming the attacker can start its own program on the same core as the model, the attacker can time its own AMX operations in parallel with the model inference to see when AMX was used by the model. Since the OS will interleave execution of both workloads, a slow AMX operation means AMX was not used by the model recently, while a fast AMX operation implies otherwise. At the cost of the more relaxed threat model, all gadgets can be utilized with the bonus of fine-grained observations, implying fewer iterations required for secret leakage. We emphasize that this threat model does not require end-to-end timing of the MLaaS model, nor does it require access to the model logits or confidence scores, as in most already-discovered membership inference attacks.
} 







As shown in the column of the Leakage Path and Threat Model in Table~\ref{tab:gadgets}, some gadgets are exploitable remotely, while others need a local perspective, highlighting that the attack surface of \scheme{} ranges from remote services to trusted environments. Even Intel SGX enclaves running vulnerable ML code show AMX cold start timing visible to a monitoring attacker in the host OS. A privileged adversary can infer secret-dependent decisions in the enclave by timing operations. This means that OS, VM, and SGX isolation does not stop \scheme{} - the timing signal is visible to any observer capable of timing operations.

   \textcolor{black}{
The side-channel attack described in Section~\ref{sec:mycovert} operates off the assumption that a Spectre-style gadget is available in pre-existing, legitimate, non-malicious network-exposed code; we assume no attacker-controlled trojan/spy code running on the victim, modified code running on the victim, or victim collusion; this is the same threat model as Netspectre~\cite{netspectre}. 
GateBleed turns the secret-dependent speculative AMX usage into a high-resolution timing channel via the Query+Time and Timing threat models.
}

\section{Reverse Engineering}\label{sec:overview}
In this section, we describe our process for reverse engineering the AMX hardware beginning with a description of the documented AMX features, continuing with our discovery of a reuse-distance-dependent AMX latency, and concluding with a root cause analysis. 

\subsection{Intel AMX Architecture}
AMX is an on-core AI accelerator in the Intel Xeon 4th (Sapphire Rapids)
 \cite{nassif2022sapphire}, 5th~ (Emerald Rapids)~\cite{munch20242}, and 6th (Granite Rapids)~\cite{varada20252} generation scalable CPUs featuring high-throughput tile-based matrix operations such as \texttt{TDPBSSD} and \texttt{TDPBF16PS}~\cite{intel2023amx,IntelOptRefManual}. Unlike off-core accelerators, AMX instructions execute within the core pipeline.

Crucially, AMX introduces eight new 1 KB architectural \textit{tile registers} capable of holding $16 \times 64$ \texttt{int8} or $16 \times 32$ \texttt{bfloat16} matrices~\cite{intel2023amx}, resulting in a total tile storage overhead of 8 KB, one-fourth of a 32 KB L1 cache. Due to this, AMX achieves up to 1024 int or 512 bfloat16 FMA operations per cycle, outperforming AVX-512~\cite{Intel_Arc_Ins_Set_Extensions}. 

Before execution, AMX tiles must be configured using the \texttt{LDTILECFG} instruction to specify the number of rows (up to 16), bytes per row (number of columns), and starting row~\cite{kalamkar2019study}. 
Data is transferred into tiles using \texttt{TILELOADD}/\texttt{TILELOADDT1}. The matrix multiplication instructions are \texttt{TDPBF16PS} for \texttt{bfloat16} input and \texttt{TDPBSSD}, \texttt{TDPBSUD}, \texttt{TDPBUSD}, and \texttt{TDPBUUD} for signed and unsigned 8-bit integers~\cite{IntelOptRefManual}. After computation, tiles are stored back into memory via \texttt{TILESTORED}.
\texttt{TILEZERO} zeroes tile registers, and \texttt{TILERELEASE} deactivates AMX, minimizing context switch overhead by avoiding the need to save 8 KB of tile register data.

 \subsection{Power-Gating,  Latency \& Privacy Leakage}\label{subsec:overview}


In this experiment, we measured how long it takes to execute a single AMX multiplication instruction while varying the intervals between consecutive executions. 
Figure~\ref{fig:Performance_Stages} shows that the latency of a \texttt{TDPBSSD} (signed-signed 8-bit integer matrix multiplication) differs depending on the time since the AMX unit was last used; that is, \textit{reuse distance}. 
By changing the length of these intervals, we observed five distinct execution times for the AMX multiplication instruction.

We noticed that this latency goes through five distinct \textit{performance stages} which correspond to a \textcolor{black}{50, 600, 6000, 9000, and 20,000 CPU cycle latency to perform a \texttt{TDPBSSD}}. We classified these into performance states, with the shortest execution time labeled as the \textit{Warm State} and the longer execution times labeled as \textit{Cold States} - specifically Cold State 1 (the second shortest), Cold State 2, Cold State 3, and Cold State 4 (the longest). This is the foundation of our \scheme{} attack, enabling both covert and side-channel attacks.  In the next section, we reverse engineer the root cause of this behavior and show that these stages constitute a class of timing leakage rooted entirely in on-core AMX accelerator power management.



 \definecolor{myblue}{HTML}{007799}
\definecolor{myorange}{HTML}{FF8888}

\begin{figure}[!bp]
    \centering

    \includegraphics[width=0.7\columnwidth]{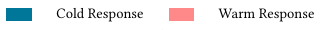}

    \noindent
    \begin{minipage}[t]{0.49\columnwidth}
        \centering
        \includegraphics[width=\linewidth]{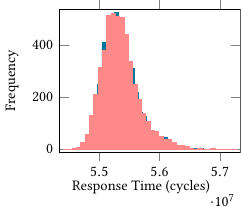}
        {\footnotesize (a) Non Member }
    \end{minipage}%
    \hfill
    \begin{minipage}[t]{0.49\columnwidth}
        \centering
        \includegraphics[width=\linewidth]{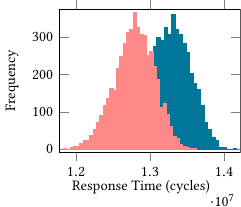}
        {\footnotesize (b) Member }
    \end{minipage}

    \caption{The response time distributions of an end-to-end \textit{transformer} model inference.}
    \label{fig:response_comparison}
\end{figure}

\begin{figure}[!htbp]
    \centering
    \includegraphics[width=0.9\linewidth]{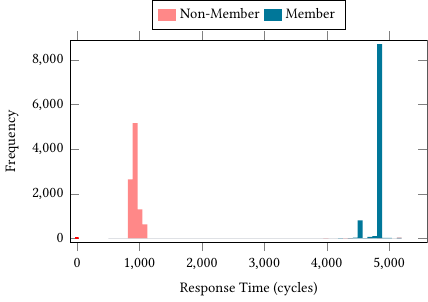}

    \caption{\textcolor{black}{AMX Operation Latency Cross-Process after CNN Inference for Member vs. Non Member.}}
    \label{fig:AMX-time_diff}
\end{figure}

\textcolor{black}{\scheme{} exploit the timing difference created by Intel AMX power gating cross process and remote with the knowledge of the library (for example having the routing logic in MoE or early exit in MSDNet shown in Table~\ref{tab:gadgets}), 
to leak secrets such as the membership of the input or private parameters of a deployed model. Figure~\ref{fig:AMX-time_diff} shows that the TMUL latency is over 4000 CPU cycles different when executed after a member inference in CNN or a non-member inference, enabling a cross-process \scheme{} attack on CNNs. We show that the leakage remains resilient even under moderate scheduling interference, with timing margins exceeding 4000 cycles between member and non-member AMX invocation illustrated in Figure~\ref{fig:AMX-time_diff}.
This phenomenon happens on the individual AMX TMUL operations due to power gating (we discuss the root cause in the next section), but we can see the effect cascading over the entire block of AMX instructions, even in an end-to-end large-scale Transformer. For example Figure~\ref{fig:response_comparison}  compares (a) non-member 
    (b) member 
    resulting in a 500,000  CPU cycle timing difference in an end-to-end inference of a transformer model, depending on the membership status of the input, enabling a remote MIA attack on the Transformer Model.  }



\subsection{Root Cause Analysis}\label{sec:reverseeng}

To understand what novel capabilities \scheme{} provide to the attacker and what privileges are required for this exploit, as well as how to mitigate it, we systematically investigated the root cause of Intel AMX performance stages and confirm that the root cause is power-state transitions driven by AMX’s independent power management, rather than core frequency scaling, operand dependencies, SMT, value dependency, SGX, or any of the core power savings settings; that is, C-states and C1E. 

The following is a summary of the main findings.

\subsubsection*{\textbf{Frequency Scaling and Throttling Effect}}
\begin{figure}[!htbp]
    \centering
\includegraphics[width=0.5\textwidth]{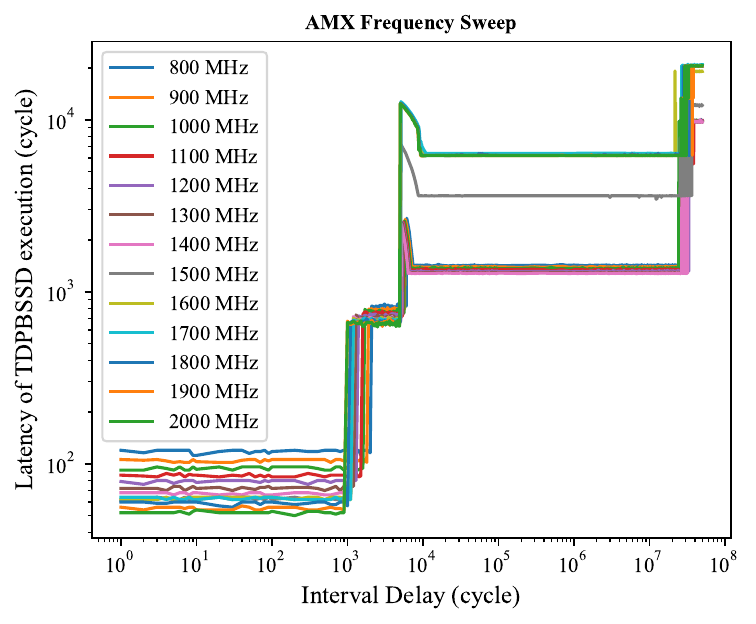}
    \caption{\scheme{} leakage under various fixed core frequencies. 
    }
    \label{fig:freqsweep}
\end{figure}

Frequency scaling and the Intel Turbo Boost feature are the root cause of many software covert channels, such as ~\cite{wang2022hertzbleed,Lipp2021Platypus, liu2022frequency, kogler2023collide+} or even physical side channels~\cite{wei2018know,Batina2019USENIX}. 
However, disabling Turbo Boost did not have an impact on AMX performance stages, which this attack exploits, confirming that CPU frequency scaling (DVFS) is not responsible for observed timing variations. Fixing the CPU frequency showed that while timing differences changed with frequency, even at 800 MHz, a 9,000-cycle gap remained exploitable. Figure~\ref{fig:freqsweep}  shows a frequency and wait-time delay sweep, observing \texttt{TDPBSSD} timings, which showcases the same five-stage performance trend across all core frequencies. 

Thus, the attacker does not require the privilege of Turbo Boost or the ability to set a fixed frequency on the victim's CPU for the attack to work. 
In addition, disabling Turbo Boost or frequency locking does not mitigate \scheme, unlike Hertzbleed~\cite{wang2022hertzbleed, liu2022frequency}. This significantly expands attackers' capability over frequency throttling-based covert channels\cite{wang2022hertzbleed, liu2022frequency}.

\subsubsection*{\textbf{Core C-states}}

C-states are CPU power-saving modes. 
C-states are numbered as follows:
C0 – The core is fully active and executes instructions.
C1 – The core is idle, but can return to C0 quickly.
C2, C3, …, Cn – Deeper sleep states; progressively more parts of the core are turned off.
Each deeper state saves more power, but takes longer to wake up. Although great for efficiency, they can unintentionally leak information through timing side channels when switching between power states - a root cause exploited by IdleLeak~\cite{rauscher2024idleleak}. We enabled and disabled the C-state feature in the system BIOS and observed that the performance stages remain unchanged; see table~\ref{tab:configurations}.  CPU C-states are not the root cause.

\subsubsection*{\textbf{C1E, Busy Waiting \& usleep}}

We check if CPU C states can affect AMX execution times via the \texttt{sleep} function, a function that explicitly puts the CPU in C1 for short waits and C6 for long waits.
Using \texttt{sleep} instead of busy waiting (to activate C states C1 and C6), if we disable enhanced C1 (C1E) while C states are enabled, cold stage 4 remains visible at a 20,000-cycle latency. C1E is an "Enhanced Idle" CPU state that combines clock gating and voltage reduction.
It is deeper than C1, but still has low latency.
With C states and C1E both enabled, we observe that cold stage 4 induces a 9,000-cycle latency while cold stage 2 induces a 3,000-cycle latency, indicating that while conventional C states do not affect AMX power gating, C1E does: C1E prevents AMX from entering deepest sleep, but it does not remove all exploitable stages. Thus, C1E is not a cause itself and therefore, disabling C1E or C-state does not mitigate \scheme.

\subsubsection*{\textbf{Value Dependency}}

We varied the operand value and noticed that this
does not affect the five performance stages,  
ruling out value dependence as the root cause of five AMX performance modes.

\subsubsection*{\textbf{Power Limit}}

Some covert channels are limited to only the lower power limits of the CPU~\cite{liu2022frequency, wang2022hertzbleed, Lipp2021Platypus}. Thus, we varied the power limit from the full range of possible power limits (126.0-454.0 watts) in the PKG domain to see if \scheme{} is limited to a certain power limit.  
All stages were present in all power limits. Thus, power limits are not the root cause of the observed behavior in AMX stages, and changing them would not mitigate \scheme, unlike ~\cite{liu2022frequency, wang2022hertzbleed, Lipp2021Platypus}.

\subsubsection*{\textbf{Prefetching Effect}}

We hypothesized that the latency stage behavior is due to differing cache hits/misses incurred when loading a tile register with the \texttt{TILELOADD} command. However, we observe this latency stage behavior for AMX instructions that do not touch the cache, such as \texttt{TDPBSSD}. Therefore, hardware prefetching is not the cause.

\subsubsection*{\textbf{Kernel handling}}

We tested both RHEL 9.4 and Ubuntu 22.04 OS. All five performance stages in AMX exist in both versions.

\subsubsection*{\textbf{Multi threading}}

Our reverse engineering shows that AMX is not shared among threads. \scheme{} is not exploiting contention among threads, unlike other covert channels~\cite{zhao2022binoculars, gras_translation_2018}
and thus cannot be mitigated by turning off multithreading/SMT.

\subsubsection*{\textbf{Power Consumption}}

\begin{figure}[!htbp]
    \centering
    \includegraphics[width=\linewidth]{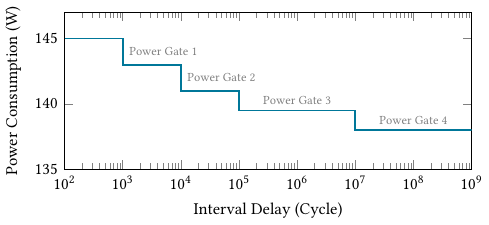}



    \caption{AMX power consumption clearly showing sharp, stepwise power gating transitions at defined interval delays.}
    \label{fig:power_gating_steps}
\end{figure}

Finally, to check power consumption in the different states, we implement a workload that performs a \texttt{TDPBSSD} and then waits for the minimal amount of time to keep AMX in a particular stage, and another identical workload in which the \texttt{TDPBSSD} instruction is omitted. We run the workloads on every core and gather the average power consumption over a period of 10 seconds. By comparing the power consumption of the first workload with the power consumption of the second workload, we can isolate the contribution of AMX in the power stages. Figure~\ref{fig:power_gating_steps}
shows the wattages we obtained along with the percent increase.

\definecolor{myblue}{HTML}{007799}
\definecolor{myorange}{HTML}{FF8888}

The gradual decrease from Stage 0 (142.08W) to Stage 4 (138.49W) aligns with staged power gating, where AMX transitions through intermediate power states before full gating.

 The state transitions induce latency shifts ranging from 50 to over 20,000 cycles, and corresponding changes in package-level power consumption, confirming the presence of an undocumented, staged power gating. Furthermore, we observed these effects even inside Intel SGX enclaves, indicating that AMX power residency is not confined by enclave or OS-level privilege boundaries.

Therefore, we attribute the root cause of \scheme{} to a novel form of \emph{unprivileged, AMX-local power gating}—a distinct microarchitectural mechanism not captured by any previously documented covert-channel primitive. This design-level behavior bypasses defenses targeting traditional timing channels (e.g., cache, TLB, SMT, or DVFS) and unlocks fundamentally new capabilities for attackers inherent in Intel AMX hardware implementation. These capabilities include bypassing defenses designed for cache and TLB-based covert channels, circumventing DVFS-based attack defenses, overcoming noise-based detection differences, evading detection due to high-magnitude timing leakage, achieving single-instruction activation, and operating securely within contexts such as Intel SGX enclaves. 

\section{\scheme{} Attack}
\label{sec:attack}
We present the \scheme{} attack, its building blocks, and \scheme{} gadgets. We identify real-world examples of \scheme{} gadgets in Table~\ref{tab:gadgets}.

\subsection{Attack Building Blocks}
Intel AMX accelerates both training and inference workloads across AI applications. Given AMX's performance profile, most modern neural networks, Transformers, GNNs, expert models, and early-exit CNNs routinely dispatch heavy matrix multiplications (matmuls) through AMX hardware. If these matrix operations are triggered conditionally based on a secret—such as token routing in a mix of experts (MoE) model, prediction confidence thresholds in an early exit model, or key presence in a KV cache—the hardware produces timing differences that correlate with the internal model state.

We define a \textit{\scheme{} Gadget} as an execution path in code that results in the triggering of Intel AMX instructions (e.g., \texttt{TDPBSSD}) based on a sensitive, private, or input-dependent decision variable. 


It follows a three-phase sequence:

   \noindent\textbf{(1)}\textbf{ Reset phase: }
    Ensures that AMX is in a lower power state. 
   
  \noindent \textbf{(2)} \textbf{Trigger Phase: }The Victim ML 
    conditionally executes an AMX instruction based on the private value.
        \begin{itemize}
            \item If \texttt{secret\_bit = 1}, execute an AMX operation from a cold state, inducing a high-latency transition.
            \item If \texttt{secret\_bit = 0}, either skip AMX or execute from a warm state, causing minimal delay.
        \end{itemize}
    
    \noindent\textbf{(3)} \textbf{Measure phase:} The attacker 
    measures the response time of the sender. If low, the receiver infers a 0; otherwise, 1. 


\subsection{Generic Magnifier}
\scheme{} also introduces a single-instruction magnification gadget that amplifies subtle microarchitectural timing differences into measurable delays, even under coarse timing conditions (e.g., 5\,\textmu s granularity in Chrome’s \texttt{performance.now()}). This enables attacks in restricted environments such as browsers, edge virtual machines, or WebAssembly runtimes, where high-resolution timers and privileged instructions are not available. Intel AMX is not currently accessible from browser sandboxes, but we anticipate that as ML moves to the edge, accelerators will be accessible from these restricted environments. 

Unlike 
Hacky Racers~\cite{HackyRacers2023ASPLOS}, which requires long instruction streams and exhibit a high microarchitectural footprint detectable by side-channel defense mechanisms, our approach leverages the reuse-distance-dependent latency of Intel AMX matrix multiplication instructions. Specifically, when the AMX unit is power-gated after a period of inactivity, a subsequent instruction such as \texttt{TDPBSSD} incurs a latency penalty of up to 20{,}000 cycles. We exploit this behavior by aligning the timing of an AMX instruction just before the power gate such that even a minor perturbation (e.g., from a cache miss or instruction port contention) can tip the unit into a colder power state, causing a sharp latency increase.

\begin{figure}[!htbp]
\centering\includegraphics[width=0.69\linewidth]{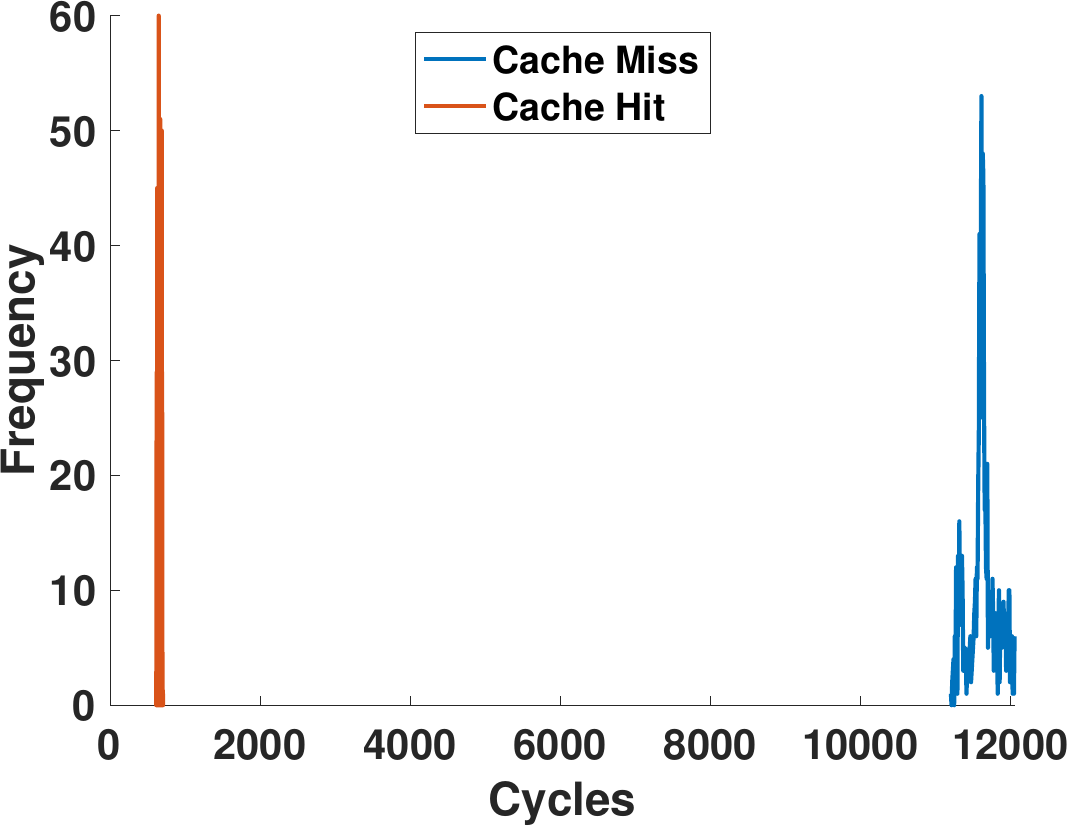}
\caption{Timing amplification using a single AMX instruction: a 200-cycle cache miss is magnified to an \textasciitilde11{,}000-cycle timing gap, bypassing timer resolution coarsening defenses.
}
\label{fig:magnify-flush-reload}
\end{figure}

This setup turns otherwise unobservable microarchitectural delays (on the order of 100–200 cycles) into coarse-timer-visible effects exceeding 11{,}000 cycles. This effect is illustrated in Figure~\ref{fig:magnify-flush-reload}, where a typical cache hit/miss delay is amplified into a \textasciitilde5.5\,\textmu s timing gap, defeating deployed timer coarsening defenses in real-world web browsers.

The gadget requires only a single \texttt{TDPBSSD} invocation and avoids any speculative execution, memory flushing, or branching behavior, rendering it nearly invisible to current microarchitectural attack detectors. Its simplicity and low-profile execution pattern make it suitable for chaining with any reuse-sensitive instruction or timing channel, including cache access, port contention, and instruction ordering. 

\subsection{Exploitable Benign Gadgets} \label{sec:gadgdetail}
 All listed gadgets share the same fundamental leakage mechanism. As described in Section~\ref{subsec:overview}, when an AMX instruction is invoked from a cold (power-gated) state, it incurs a substantial warm-up latency. All gadgets exploit this reuse-distance-driven latency effect: the first AMX matmul on a given execution path will be dramatically slower if the unit was idle beforehand. By making AMX invocation conditional on a secret or private branch, the timing of the code becomes correlated with the secret. Notably, this timing difference manifests even if both branches perform nominally identical workloads. For instance, even if a model tries to execute all experts in an MoE layer to avoid branching, the first semantically non-noop matmul still triggers a cold-start penalty, leaking which expert was needed. Similarly, padding or dummy computing an 'early exit' does not eliminate the initial delay when the AMX unit switches from idle to active. In every gadget, the same pattern holds: A secret-gated AMX operation creates a timing fingerprint governed by the accelerator's previous usage history (that is, the reuse distance). This uniform root cause gives us confidence that any such conditional-AMX code path can exhibit \scheme{} leakage.

\textcolor{black}{We identify \scheme{} gadgets by searching ML libraries for the existence of branches leading to a matrix operation. The matrix multiplication can be compiled to be optimized with Intel AMX.}
We grouped identified gadgets into three classes:  
(1) \textit{Token-routing branches} (e.g., MoE experts, tool dispatch);  
(2) \textit{Confidence-gated exits} (e.g., early-exit classifiers); and  
(3) \textit{Session/context-sensitive toggles} (e.g., KV-cache, quantization paths).  
Only classes (1) and (2) are input-dependent and security-sensitive; class (3) enables attacker-agnostic fingerprinting. Notably, all trigger real AMX matmuls.

These are not speculative code samples, but production-grade code where secret-dependent variables (e.g., confidence scores, routing decisions, session flags) conditionally guard high-throughput matrix computations optimized with AMX backends such as oneDNN and MLAS. In each case, a measurable power-gated latency difference emerges from the first semantically meaningful AMX instruction, even under balanced control flow.


\scheme{}-like leakage is not a mere theoretical construct but a plausible risk across a wide range of ML libraries and models that employ conditional high-performance routines. Indeed, our investigation uncovered more than a dozen potential gadgets \scheme{} in production ML frameworks. 
As Table~\ref{tab:gadgets} shows, these gadgets span real workloads in NLP (e.g., Mixtral), GNNs (e.g., RGATConv), vision (e.g., SkipNet, MSDNet), agent-based LLM frameworks (LangChain, AutoGen), and ML inference APIs (OpenAI Function Calling)
across widely-used ML frameworks (e.g., Hugging Face, PyTorch, TensorFlow, ONNX Runtime, DeepSpeed).



\begin{table}[!htbp]
    \centering
    \resizebox{\columnwidth}{!}{%
    \begin{tabular}{l|c|c|c|c}
        \textbf{Attribute} & \textbf{Value} & \textbf{GB} & \textbf{Hz} & \textbf{IL} \\
        \hline
        P-state control & Autonomous (hardware-only) & \checkmark & \checkmark & \checkmark \\
        P-state control & Legacy (OS-only) & \checkmark & $\circ$ & \checkmark \\
        P-state control & Cooperative & \checkmark & \checkmark & \checkmark \\
        P-state control & Disabled & \checkmark & $\circ$ & \checkmark \\
        OS & RHEL 9.4 & \checkmark & \checkmark & \checkmark\\
        OS & RHEL 9.5 & $\circ$ & \checkmark & \checkmark \\
        OS & Ubuntu 22.04 & \checkmark & \checkmark & \checkmark \\
        UEFI Version & SRV650-v3-3.14 (May 2024) & \checkmark & \checkmark & \checkmark \\
        UEFI Version & SRV650-v3-3.20 (June 2024) & $\circ$ & \checkmark & \checkmark \\
        Platform Power & Minimal Power & \checkmark & \checkmark & \checkmark \\
        Platform Power & Maximum Performance & \checkmark & \checkmark & \checkmark \\
        Platform Power & Efficiency, Favor Power & \checkmark & \checkmark & \checkmark \\
        Platform Power & Efficiency, Favor Performance & \checkmark &  \checkmark & \checkmark \\
        Turbo Boost & Enabled & \checkmark & \checkmark & \checkmark \\
        Turbo Boost & Disabled & \checkmark & \textbf{x} & \checkmark \\
        All prefetchers & Disabled & \checkmark & \checkmark & \checkmark \\
        C-States & Enabled & \checkmark & \checkmark & \checkmark \\
        C-States & Disabled & \checkmark & \checkmark & \textbf{x}\\
        C1E & Enabled & \checkmark & \checkmark & \checkmark \\
        C1E & Disabled & \checkmark & \checkmark & \checkmark \\
    \end{tabular}%
    }
    \caption{Configuration settings for \scheme, Hertzbleed~\cite{wang2022hertzbleed}, and IdleLeak~\cite{rauscher2024idleleak} across various system configurations. GB, Hz, and IL refer to \textit{\scheme}, \textit{Hertzbleed}, and \textit{IdleLeak}, respectively.}
    \label{tab:configurations}
\end{table}

\section{Results}\label{sec:results}
This section presents experimental settings for ~\scheme{} results as a side-channel attack against ML models via AMX gadgets found in real-world codebases. To our knowledge, this is the first side-channel attack on machine learning privacy utilizing hardware acceleration. We then present~\scheme{} results as a novel passive side channel with exceptionally high bandwidth. We finally present~\scheme{} as a generic magnifier capable of amplifying subtle microarchitectural timing differences into visible differences in timer-constrained environments. 

\subsection{Experimental Setting}\label{sec:setting}

Our investigations utilized a server as a victim running {Red Hat Enterprise Linux 9.4} with Linux Kernel 5.14, powered by
an Intel Xeon Gold 5420+ CPU of the Sapphire Rapids microarchitecture. 
The network we used is a production network with an average daily traffic of tens of terabytes, employing no network isolation. 
The attacker/client is a Skylake desktop in remote settings. Table~\ref{tab:configurations} summarizes the OS and UEFI settings we tested, along with how they affect the operations of two state-of-the-art side-channel attacks: Hertzbleed~\cite{wang2022hertzbleed} and IdleLeak~\cite{rauscher2024idleleak}. 

\textcolor{black}{Table~\ref{tab:gadgets} catalogs a broad set of AMX-triggering gadgets. 
In this work, we implement realistic AMX-based PoCs for selected gadgets that represent each gadget class. 
These include Mixtral (HF), TensorFlow MoE, DeepSpeed MoE, ONNX Runtime MoE, Mixtral (llama.cpp) for expert routing, and BranchyNet/MSDNet-style early exit CNNs and transformers for confidence-based control. 
}

\subsection{Leaking Routing Decisions in Mixture of Transformer Experts (MoEs) }\label{sec:MoE}

\textcolor{black}{To reflect real-world deployments, our PoC implements a heterogeneous Mixture-of-Experts model with Intel AMX patterned directly after \textit{Mixtral (HF)}, as listed in Table~\ref{tab:gadgets}. Like Mixtral, our model activates a subset of experts per token (one out of two in our case), with AMX-accelerated matrix multiplications executed only for the selected expert. The higher-capacity expert matches the configuration in Table~\ref{tab:transformerspecs}, while the lower-capacity expert has a reduced Transformer depth. This mirrors Mixtral’s expert asymmetry and sparse dispatch behavior. We used a training set of 784 English sentences and a test set size of 300 English sentences to train the model.}



\begin{table}[!htbp]
\centering
\renewcommand{\arraystretch}{1.2}
\resizebox{0.48\textwidth}{!}{%
\begin{tabular}{|l|c|c|}
\hline
\multirow{2}{*}{\textbf{Parameter}} & \multicolumn{2}{c|}{\textbf{Expert}} \\
                                   & \textbf{1 (High Capacity)} & \textbf{2 (Low Capacity)} \\
\hline
Hidden size           & 256     & 256 \\
Intermediate size     & 256     & 256 \\
Number of heads       & 4       & 4 \\
Attention type        & Multi-headed & Multi-headed \\
Embedding size        & 256     & 256 \\
Number of layers      & 24      & \textbf{10–22 (varied)} \\
Dropout               & 0.1     & 0.1 \\
Activation            & ReLU    & ReLU \\
Parameter sharing     & None    & None \\
\hline
\end{tabular}%
}
\caption{Transformer Specifications in Heterogeneous MoE.}
\label{tab:transformerspecs}
\end{table}

\begin{figure}[!htbp]
\centering
\includegraphics[width=0.5\textwidth]{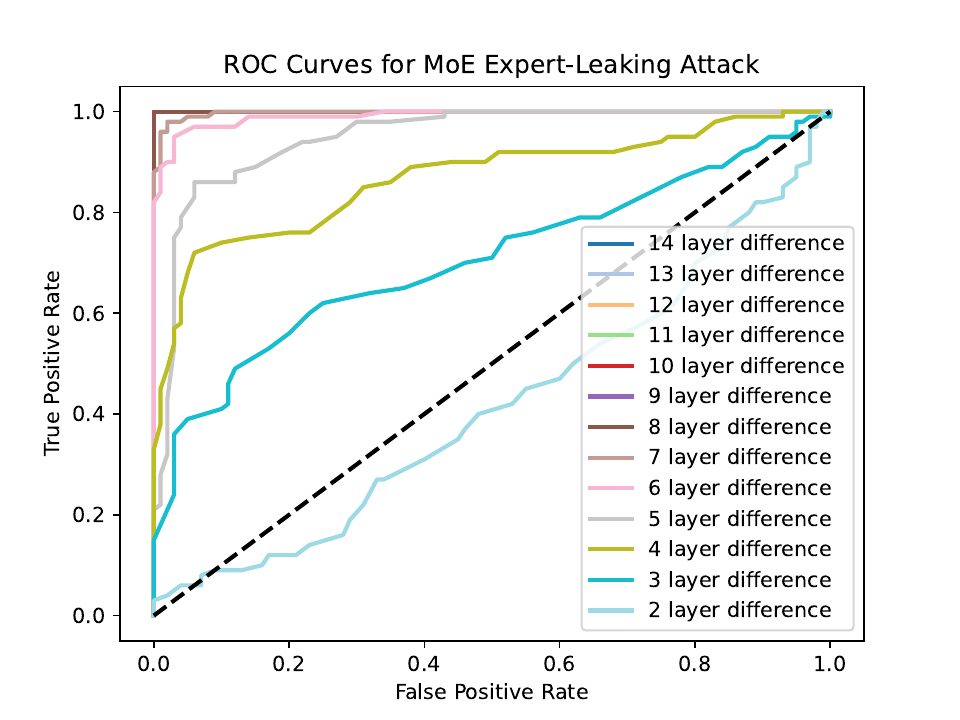}
\caption{\textcolor{black}{ROC for \scheme{} attack on MoEs. 8-layer differences up to 14-layer differences have 100\% accurate classification, hence they overlap over the pink curve in the plot. 
}}
\label{fig:moe-roc}
\end{figure}

\textcolor{black}{\scheme{} successfully infers the expert routing index in a heterogeneous Mixture-of-Experts Transformer with an overall accuracy of \textbf{100\%},
indicating it reliably distinguishes between which expert was activated even when model parameters and logits are hidden. 
We perform a comprehensive sensitivity study: Figure~\ref{fig:moe-roc} presents ROC curves for varying differences in expert depth. When the lower-capacity expert has 10–16 layers and the higher-capacity expert remains fixed at 24 layers (i.e., a layer gap of 8 or more), \scheme{} achieves perfect separation: 100\% true positive and true negative rates, with zero false positives or false negatives. 
As the layer gap narrows, leakage weakens but remains significant. Table~\ref{tab:attack_metrics} compares the success rate, FP, FN, TP, and TN rates numerically.}

\begin{table*}[!htbp]
\centering
\begin{tabular}{@{}lcccccc@{}}
\toprule
\textbf{Attack} & \textbf{Accuracy / Success} & \textbf{TPR} & \textbf{FPR} & \textbf{TNR} & \textbf{FNR} & \textbf{Precision} \\
\midrule
MoE (Layer Gap $\geq$ 8)    & 100\% & 100\% & 0\%  & 100\% & 0\%  & 1.0 \\
MoE (Layer Gap = 7)    & 98\%  & 98\%  & 2\%  & 98\%  & 2\%  & 0.98 \\
MoE (Layer Gap = 6)    & 96\%  & 95\%  & 3\%  & 97\%  & 5\%  & 0.97 \\
MoE (Layer Gap = 5)    & 90\%  & 86\%  & 6\%  & 94\%  & 14\% & 0.93 \\
MoE (Layer Gap = 4)    & 82\%  & 74\%  & 10\% & 90\%  & 26\% & 0.88 \\
MoE (Layer Gap = 3)    & 69\%  & 62\%  & 25\% & 75\%  & 38\% & 0.71 \\
MoE (Layer Gap = 2)    & 46\%  & 40\%  & 48\% & 52\%  & 60\% & 0.45 \\
\midrule
Early Exit CNN 
& 99.72\%         & 99.99\%     & 0.54\%     & 99.46\%         & 0.01\%         & 0.99          \\
Early Exit Transformer & 100\% & 100\% & 0\% & 100\% & 0\% & 1.0 \\
Transformer MIA         & 81\%            & 78\%        & 16\%       & 84\%       & 22\%       & 0.89          \\
\bottomrule
\end{tabular}
\caption{Evaluation metrics across verified categories of end-to-end attacks  with \scheme{} timing/AMX usage leakage.}
\label{tab:attack_metrics}
\end{table*}

\subsection{Leaking Early-Exit Decisions \& Membership via AMX Timing}


\textcolor{black}{Our PoC targets an early-exit convolutional neural network (CNN) following the structure of BranchyNet~\cite{teerapittayanon2016branchynet} described in Table~\ref{tab:gadgets}. The model contains six layers: a convolution followed by max-pooling and ReLU, then two fully connected layers. An early-exit branch is inserted after layer 2, where the model computes a softmax over logits and exits if confidence exceeds a threshold. We use a soft threshold-based condition to simulate production behavior, and the model routes either through this shallow path or the deeper full path based on this internal decision.} 





\textcolor{black}{
A special condition under which this attack takes place is that the time taken by the early exit path and full path are essentially the same, making the timing side channel ineffective.
In this attack, \scheme{} achieves a classification success rate of  99.72\% to infer whether the model exited early or executed the full path with the help of AMX power gating. The true positive rate—correctly identifying early exits—is 99.99\%, with a false positive rate of just 0.54\%. Here, we have taken the threshold as the mean of the average cycles taken by the AMX instruction following the Early Exit path and average cycles taken by the AMX instruction after the Full Path to identify the inference path. These results hold across 20{,}000 repeated trials. 
}

\textcolor{black}{To assess the sensitivity of the channel to architectural parameters, we vary the position of the early-exit condition and measure performance.}
\textcolor{black}{If sufficient confidence is computed using the softmax of the logits of the exit layer, the NN computation exits. We train and evaluate this model on the MNIST dataset.} 

\textcolor{black}{When the exit occurs after skipping four layers, the area under the ROC curve (AUC) reaches 1.0. With three skipped layers, AUC remains above 0.997. Even when only two layers are skipped, the timing remains distinguishable enough to support an AUC of 0.85, confirming that the leakage scales predictably with the compute disparity between exit paths. At one-layer difference, the signal begins to degrade, but still retains a measurable gap.}

\begin{figure}[!htbp]
    \centering \includegraphics[width=0.9\linewidth]{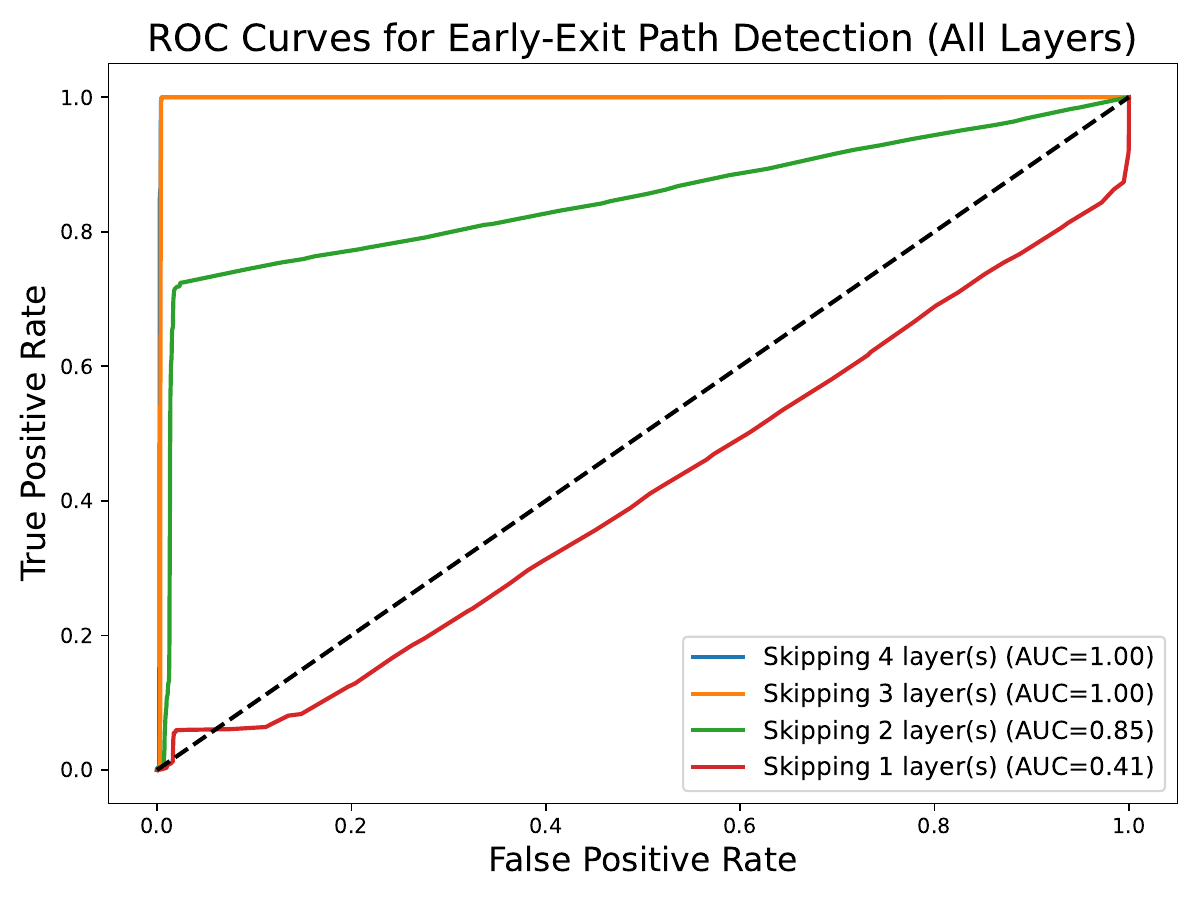}
    \caption{ROC for Early Exit CNN.}
    \label{fig:ROC-CNN}
\end{figure}



\textcolor{black}{We perform an end-to-end membership inference attack using the AMX usage signal observed by \scheme{}. For this experiment, we implemented an early-exit Transformer model to evaluate whether \scheme{} can be used to infer training data membership through timing leakage. The model consists of 24 Transformer layers and exits after layer 12 if the softmax confidence exceeds a threshold. All matrix multiplications are dispatched to Intel AMX using \texttt{TDPBSSD}, and the architecture parameters follow Table~\ref{tab:transformerspecs}.
We achieve an overall accuracy of 81\% with 78\% of members correctly identified and only 16\% of non-members misclassified. These results rival or exceed prior attacks that required full output vectors or confidence scores.}

\textcolor{black}{This is the first demonstration of a successful, end-to-end membership inference attack on an early-exit model deployed with hardware acceleration with no reliance on output vectors or model internals. The attacker requires only the ability to detect AMX usage, achievable via co-residency as discussed in Section~\ref{sec:threat}.}
\subsection{Magnification for Remote Arbitrary Address Leakage} 
\label{sec:mycovert} 
{Traditional microarchitectural side channels consist of a leakage channel and transmission channel~\cite{kiriansky18dawg}.} In this section, we show that  \scheme{} is not just a side channel targeting ML privacy—but it also serves as a highly effective transmission channel for existing microarchitectural attacks like Spectre to leak the contents of an arbitrary address, such as speculative execution vulnerabilities particularly in remote settings where prior methods fail like a realistic network.

For example, the remote Spectre attack Netspectre~\cite{netspectre} uses the power-gating optimizations in the Intel AVX-2 and AVX-512 to transmit the leaked secret over a network. However, we find that a timing channel built on a timing difference of a few hundred cycles ~\cite{wang2022hertzbleed, netspectre} is impractical  
in a realistic production network. 

The difference in execution between a fully powered and power-gated AVX-512 unit is 
about 150 cycles on Intel Xeon processors vs. 20,000 cycles for AMX. 
This two-order-of-magnitude gap in timing (AMX vs. AVX) provides a fundamentally stronger signal, which we show by comparing them as two covert channels in our scheme. 
Using an already available \scheme{} gadget in the victim code can enable remote leakage of arbitrary addresses where NetSpectre fails. This is mainly because prior microarchitectural attacks have often shown effectiveness in networks where the traffic from other users was eliminated. A real network consists of uncontrolled traffic from multiple users, services, and devices, including jitter, congestion, and firewall delays. In such environments, we show that \scheme{} was the only channel resilient enough to operate.


 

\begin{figure}[!htbp]
    \centering
    \begin{tikzpicture}
        \node[inner sep=0pt, anchor=south west] (img) 
            at (0,0) {\includegraphics[width=\linewidth]{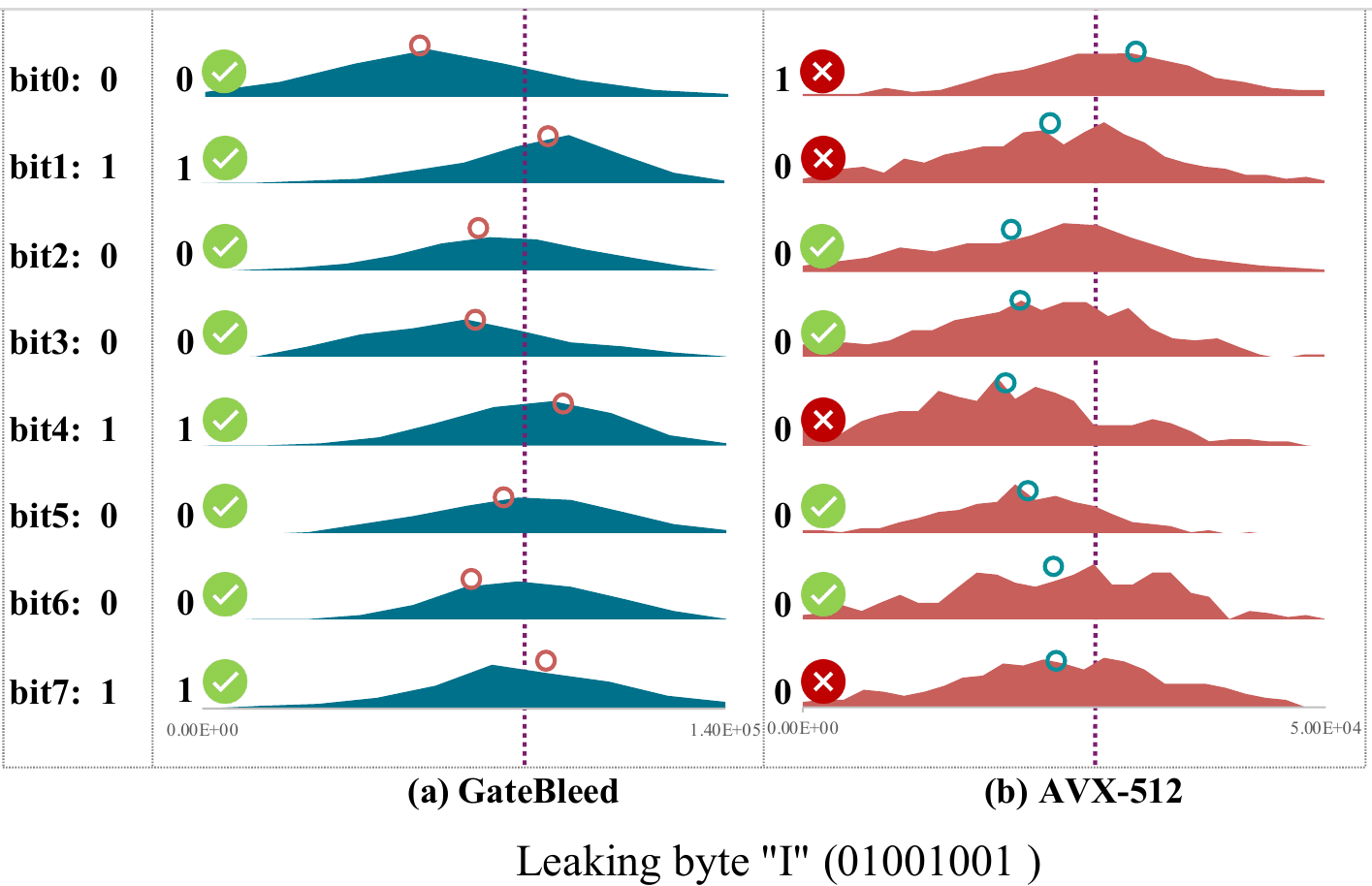}};
        
        \begin{scope}[x={(img.south east)},y={(img.north west)}]
            \node[anchor=south west] at (-0.01,0.93) 
                {\textbf{\scriptsize{\textcolor{darkblue}{ASCII of "I"}}}};
        \end{scope}
    \end{tikzpicture}
    \caption{Leaking byte \texttt{I = 01001001} over a production network. \scheme{} (left) shows separation in response-time distributions; NetSpectre with AVX-512 as transmission channel (right) fails to distinguish bits. \textcolor{black}{The x-axis is response time over the network to the attacker's and the y-axis is frequency.} Each row shows timing histograms per bit; dots show means.}
    \label{fig:leak_I_bit}
\end{figure}

Figure~\ref{fig:leak_I_bit} compares \scheme{} to AVX-512 as a covert channel on the same network. Even at 1000 trials per bit, AVX-based timing differences are fully drowned in latency noise. In contrast, \scheme{} maintains visible signal margins per bit due to AMX power-state transitions of up to 20,000 cycles. 

With a \scheme{} as a transmission channel, we leak arbitrary information across realistic network conditions by exploiting AMX warm-up latency. Our PoC demonstrates successful bit-wise recovery over a 1-hop Ethernet link at 0.07 bps (1 bit every 15 seconds) - a \textbf{70,000$\times$ improvement} over the $10^{-6}$ bps observed by AVX-512 under identical conditions. In contrast, Hertzbleed failed to leak any bits reliably across the same network, confirming that timing margins below 200 cycles collapse under real-world jitter.

In our production environment with no network isolation and tens of terabytes of daily traffic, original NetSpectre failed to reliably leak even one byte, while using \scheme{} as the transmission channel, it succeeded in transmitting 8-bit secrets with high accuracy. 
Therefore, \scheme{} as a transmission channel enables side channel attacks like Spectre to succeed in realistic remote conditions by serving as a high-bandwidth, low-noise, undetectable transmission layer.

\subsection{Noise Resilience on Real Network}
\label{subsec:noise-resilience}
\begin{figure}[!htbp]

\centering
{
\subfloat{\includegraphics[width=0.25\textwidth]{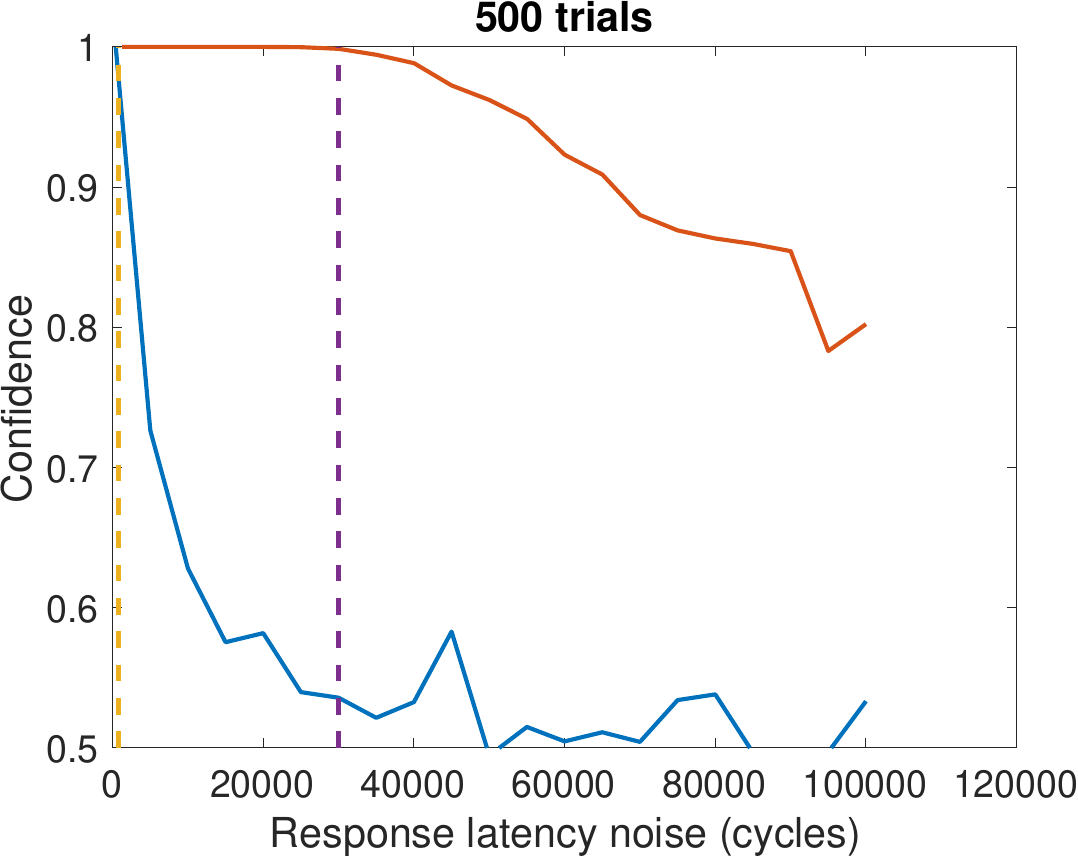}}
\subfloat{\includegraphics[width=0.25\textwidth]{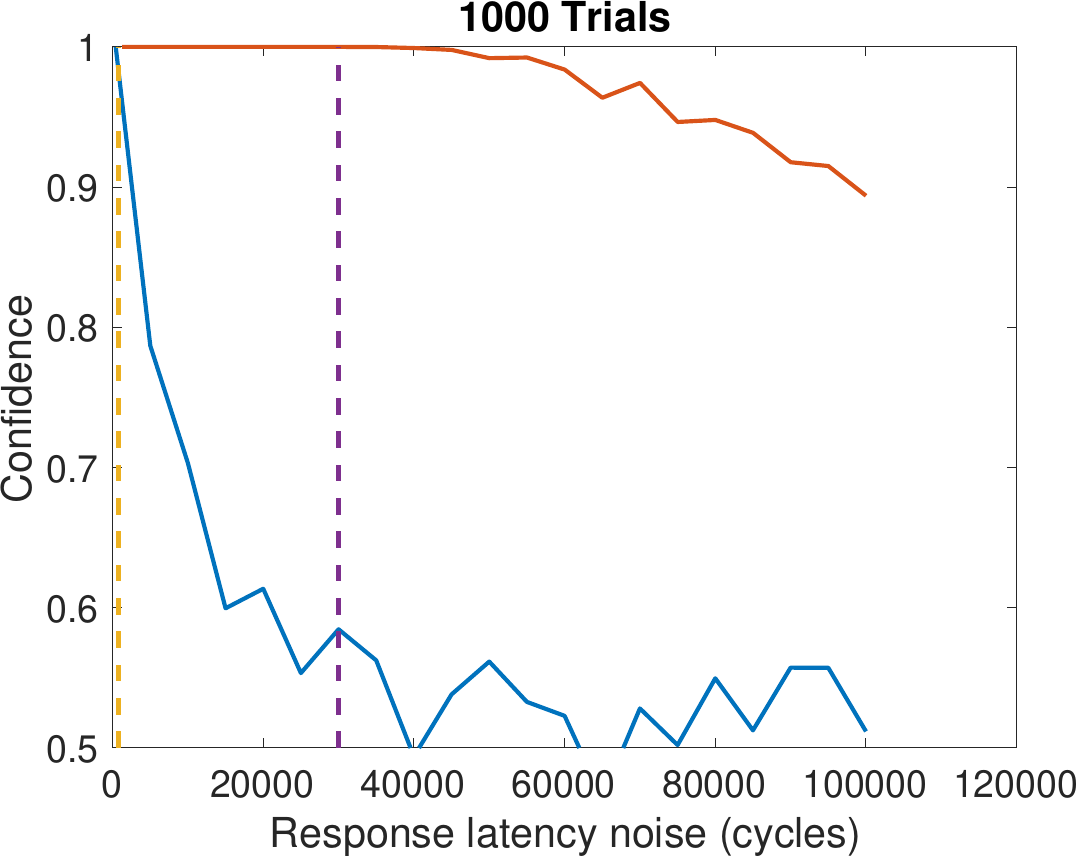}}
\caption{Noise resilience at 500 trials and 1000 trials. The orange line is \scheme, the blue line is AVX-512, the yellow dotted line is the localhost noise level, and the purple dotted line is our 1-hop noise level. 
}
\label{fig:noise-resilience}

}
   \end{figure}

\begin{figure*}
\subfloat[AVX-512 localhost]{\includegraphics[width=0.25\textwidth]{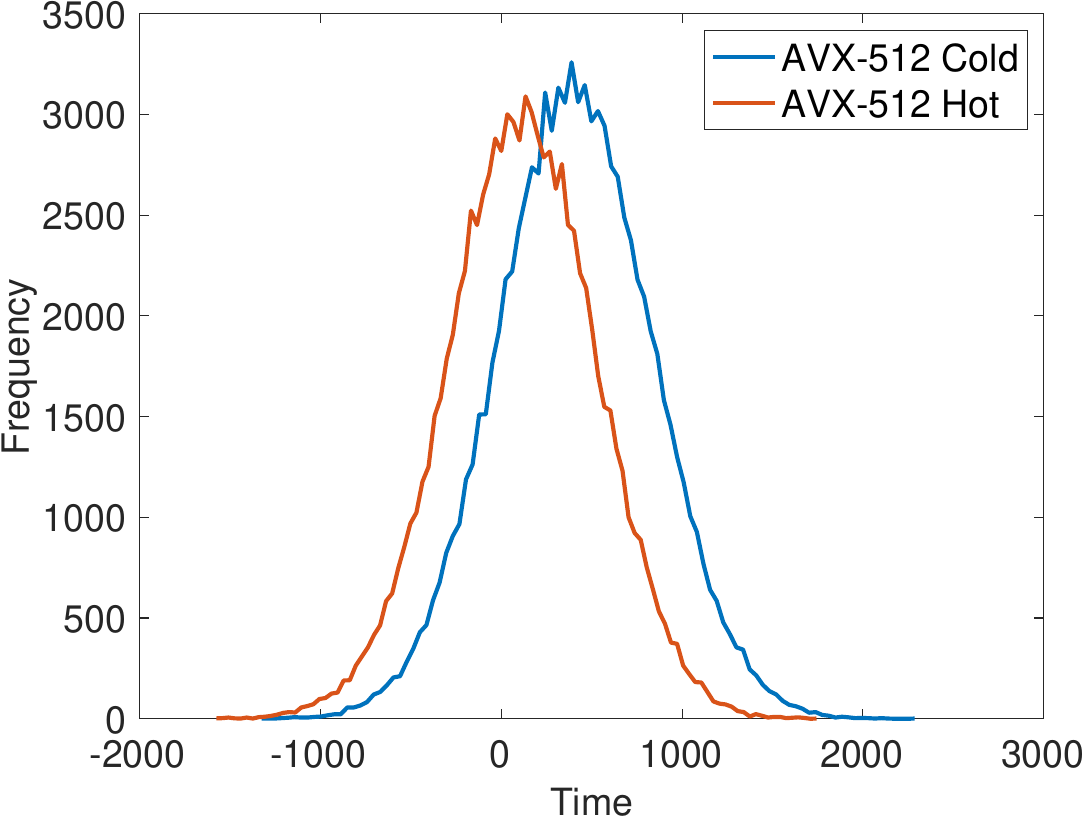}}
\subfloat[\scheme{} localhost]{\includegraphics[width=0.25\textwidth]{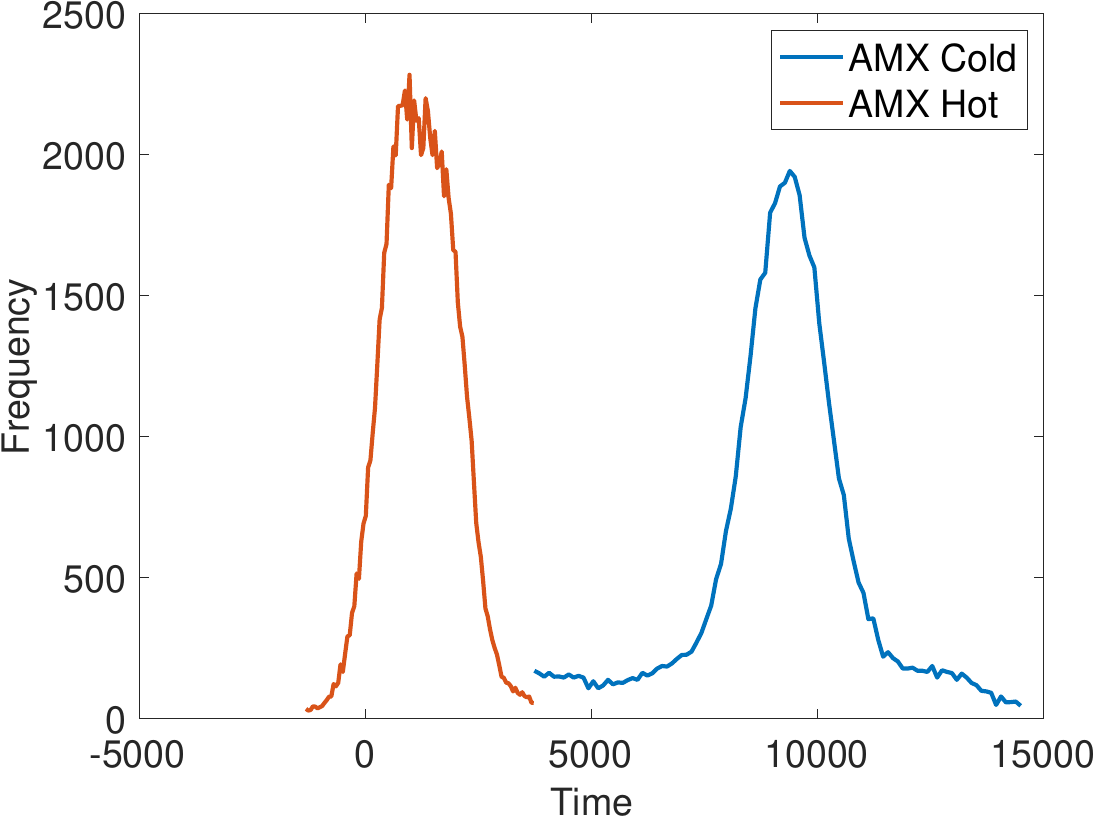}}
\subfloat[AVX-512 1 hop]{\includegraphics[width=0.25\textwidth]{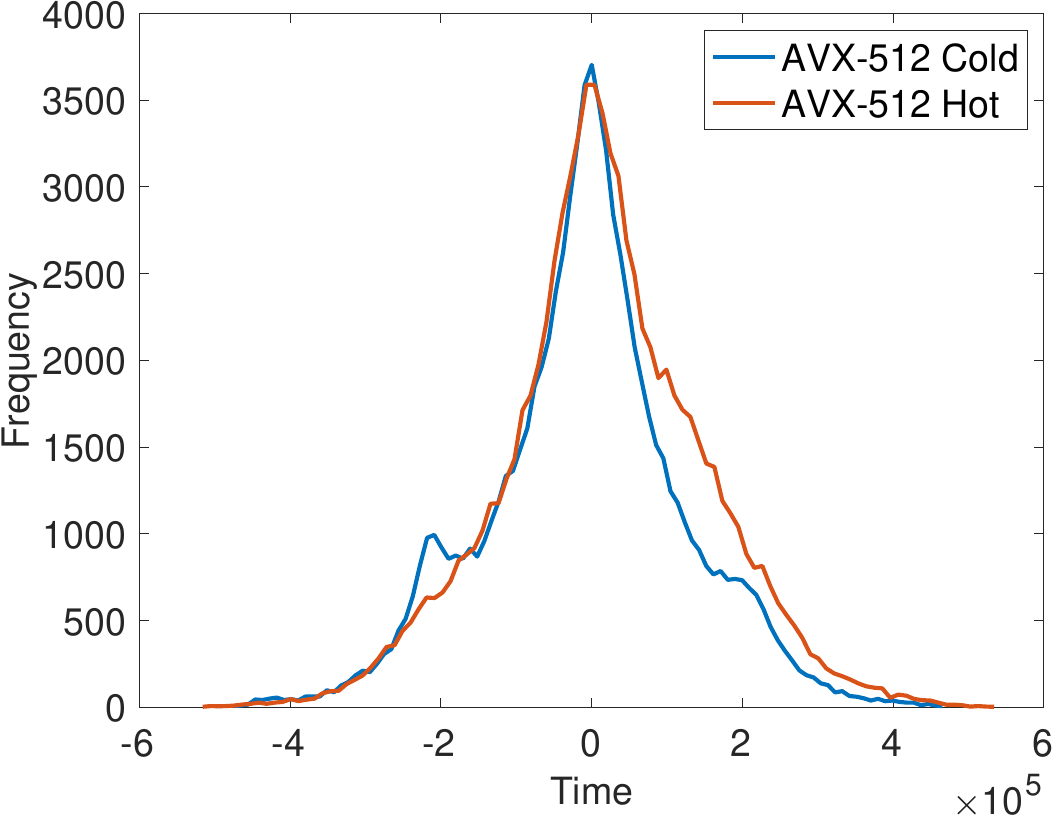}} 
\subfloat[\scheme{} 1 hop]{\includegraphics[width=0.25\textwidth]{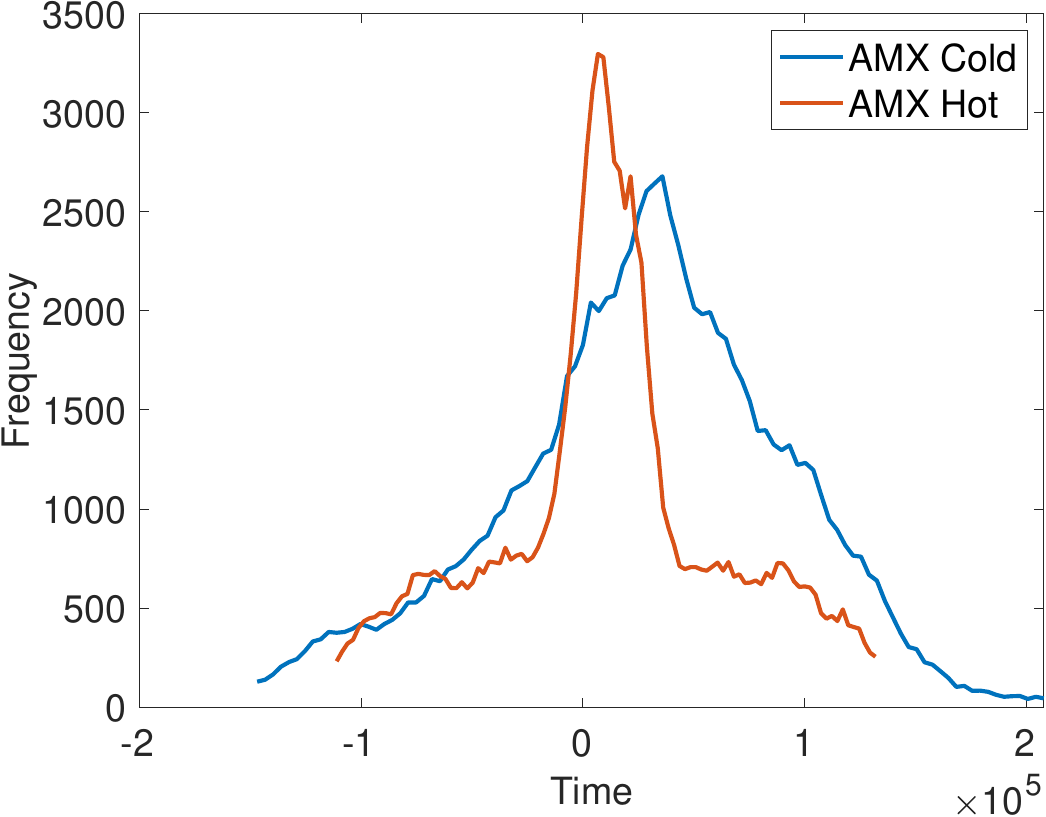}}

\caption{Comparison of local vs. remote side channel attack timing observability. 
}

\label{fig:covertchannels}
\end{figure*}

We define \textit{noise resilience} as the maximum noise level at which a covert channel can maintain 99\% confidence for a fixed trial count, with noise level equal to the variance of the network response time. We have previously seen in Figure~\ref{fig:covertchannels} that the response latencies in the networks we tested can be modeled approximately as normal distributions with different variances. In Figure~\ref{fig:noise-resilience}, we simulate network noise by applying additive white Gaussian noise of a particular power (x-axis). As network noise increases, the AVX-512 covert channel experiences a sharp decline in accuracy, approaching 0\% almost immediately. In contrast, \textsc{\scheme} maintains full resilience up to our measured 1-hop connection variance, significantly outperforming the state of the art. Note that our measured 1-hop environment had $\sigma \approx 30,000$ cycles.

Figure~\ref{fig:covertchannels} shows that exploiting the AVX-512 power gating fails on a 1-hop network connection with an entirely overlapping distribution for the secret bit $0/1$. \scheme{} leaks with high performance and stealth in both a local and a realistic production network.
   
\subsection{Timer Coarsening}
\label{subsec:timer-coarsening}


\scheme{} circumvents timer coarsening by exploiting AMX power-gating stages, which introduce latency shifts as large as 20{,}000 cycles. To suppress this channel, the timer resolution must be degraded to 10\,$\mu$s—a 20,000$\times$ coarsening over the 0.5 ns TSC in Sapphire Rapids - far beyond what is deployed in real systems (e.g., 5 $\mu$s in Chrome~\cite{chromestatusChromePlatform}).

Our results show that AVX-512 and prior side channels (e.g.,  Hertzbleed) collapse under even moderate timer coarsening and network noise. \scheme{}, by contrast, maintains a detectable signal at resolutions where others fail. Even when operating with only 500 trials, it achieves 99\% classification confidence, demonstrating that AMX's latency gap acts as a built-in timing magnifier, defeating traditional timer-based defenses.
This makes it the only channel in our evaluation that consistently achieves high-confidence leakage under production-like noise and coarse timer constraints. These results validate that power-state transitions in AMX accelerator create a far stronger and more resilient timing source than AVX and traditional microarchitectural power based effects.

  \subsection{Stealth Study}\label{subsec:mad}

\scheme{} completely eludes state-of-the-art HPC-based detection systems by leaving no observable microarchitectural footprint: no cache activity, TLB usage, or branch mispredictions. Contemporary detectors such as EVAX~\cite{EVAX2022Micro}, PerSpectron~\cite{PerSpectron}, and RHMD~\cite{RHMD2017} rely on frequent performance counter sampling to flag anomalies such as cache misses, branch mispredictions, or TLB faults. These approaches are effective against conventional covert channels, including Flush+Flush~\cite{gruss2016flush+}, Binoculars~\cite{zhao2022binoculars}, and HackyRacers~\cite{HackyRacers2023ASPLOS}, all of which inherently produce repetitive and visible side effects. In contrast, \scheme{} uses only a single AMX instruction after a passive reset phase and does not invoke any microarchitecturally anomalous instructions. With no clflush, TLB thrashing, or high-rate events, the attack resembles benign idle behavior from the detector's perspective. The only AMX-related hardware performance counter is \texttt{EXE.AMX\_BUSY} which counts the number of cycles in which Intel AMX was used; we found that including this performance counter did not improve the models' performances. 

\begin{table}[!b]
\renewcommand{\arraystretch}{1.1}
\setlength{\tabcolsep}{3.5pt}
\centering
\small
\begin{tabular}{|l|c|c|c|}
\hline
\rowcolor{gray!15}
\textbf{Attack / Gadget} & \textbf{EVAX~\cite{EVAX2022Micro}} & \textbf{PerSpectron~\cite{PerSpectron}} & \textbf{RHMD~\cite{RHMD2017}} \\
\hline
\scheme{}  & \textbf{10\%} & \textbf{9\%} & \textbf{6\%} \\
Microscope~\cite{Microscope} & 80\% & 78\% & 63\% \\
Flush+Flush~\cite{gruss2016flush+} & 99\% & 87\% & 72\% \\
Binoculars~\cite{zhao2022binoculars} & 98\% & 97\% & 85\% \\
NetSpectre~\cite{netspectre} & 97\% & 95\% & 94\% \\
Hacky Racers~\cite{HackyRacers2023ASPLOS} & 100\% & 98\% & 90\% \\
\hline
\end{tabular}
\caption{Detection accuracy of state-of-the-art detectors on known covert channels and magnifiers. \scheme{} remains undetected by all three, despite retraining.}
\label{tab:evax_perspectron_rhmd}
\end{table}

Even after extensive retraining, these detectors do not detect \scheme{} with useful accuracy. As shown in Table~\ref{tab:evax_perspectron_rhmd}, EVAX, PerSpectron, and RHMD achieve less than 10\% accuracy on \scheme{} (despite being retrained on 100 million labeled samples), while achieving more than 90\% accuracy on detectable channels. This ineffectiveness is due to the nature of \scheme{} itself: it takes advantage of the architectural latency of the AMX power-gate rather than any detectable microarchitectural side effect. The attacker merely waits for AMX to idle naturally and then issues a single matrix multiplication, producing a measurable latency gap with no suspicious footprint. This low-instruction, low-repetition channel fundamentally bypasses the pattern recognition logic of current detection techniques, rendering \scheme{} effectively invisible to today's HPC-based side-channel defenses.

\section{Countermeasures}\label{sec:defense}

Based on the root cause analysis in the section~\ref{sec:reverseeng}, methods such as disabling TurboBoost, C-states, C1E, fixing the frequency,
disabling RAPL, adding noise, or deploying cache defenses that mitigate prior attacks do not mitigate \scheme{}. Increasing the CPU's timer resolution by 20,000x is also unacceptable. Relying on state-of-the-art microarchitectural attack detectors also fails due to the high evasiveness of \scheme{}.
The root cause is not speculative execution, cache usage, or DVFS—it is a hardware-level power gating state in AMX, 
This form of leakage operates \emph{without speculation}, undermining traditional mitigations like LFENCE or retpolines~\cite{retpoline}.

Constant-time programming is a widely used defense against timing side channels on cryptographic algorithms, requiring that all code paths execute the same instructions regardless of secret inputs. This approach has been widely studied in the context of cryptographic routines, where several attacks have shown how cache-timing channels can leak secret keys even in seemingly secure implementations ~\cite{Cross-VM, 8835325, Yarom2016CacheBleedAT, 8835216, Cryptography5958048, 7163050, Curve2551944, Genkinpaper22, ConstantTime111, flush+reload}.
%

For example, in a Mixture-of-Experts (MoE) model that normally activates only 2 of 16 experts per input, constant-time enforcement requires executing \emph{all 16} experts and discarding the unused outputs. This not only increases runtime by up to $8\times$, but also fully activates the compute units (e.g., AMX) for each path, causing a sharp increase in both energy and latency.
This makes it impractical for high-throughput AI systems.

\subsection{Proposed Defenses \& Trade-offs}\label{sec:proposed}

We discuss multiple defense strategies: stage locking (always-on or fixed-state AMX), context switch-aware resetting, and hardware- or firmware-level redesigns.

\subsubsection{\textbf{Always-Warm vs. Always-Cold Trade-offs}}

In the first class of defenses, one can enforce a fixed AMX power stage throughout execution. For example, keeping AMX always warm by forcing it into the warm state eliminates any warm-up delays and thus fully masks timing variation. However, this defense imposes the highest power overhead, reaching 12\% in our measurement. 

On the other extreme, keeping AMX fully cold at Power Gate 4 yields no additional power draw, but results in the worst-case performance penalty of 35\%, as each AMX operation incurs the maximum cold-start latency of 10 $\mu$s. Intermediate fixed-stage settings offer tunable tradeoffs. 

\begin{figure*}[!t]
\centering
\includegraphics[width=\textwidth]{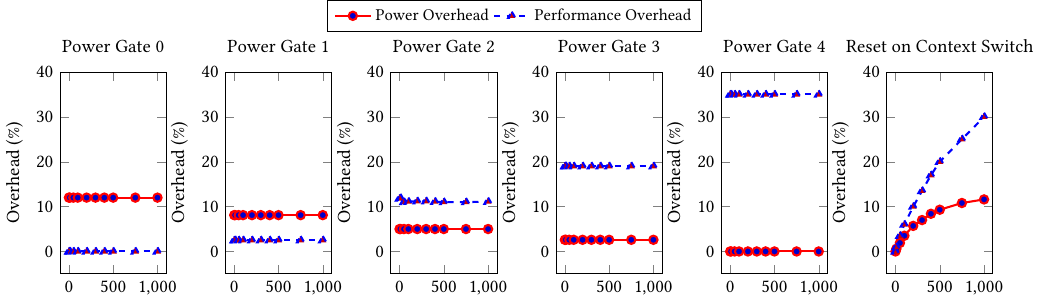}
\caption{\textcolor{black}{Comparison of AMX mitigation strategies. Each subplot shows the power (solid red) and performance (dashed blue) overhead as a function of context switch rate (per sec).}}
\label{fig:overhead}
\end{figure*}

For instance, Power Gate-1 reduces power overhead to 8.1\% while still preserving performance, with only a 2.5\% execution time increase. Power Gate-2 further lowers power usage to 5\%, though with higher latency overhead (11.1\%). This pattern demonstrates that fixed-stage defenses offer a spectrum of options with a trade-off between energy and speed. These results are shown in Figure~\ref{fig:overhead}. 

\subsubsection{\textbf{Context Switch-Aware Mitigation (OS-level)}}

A second class of defenses leverages the operating system to reset AMX state on each context switch, preventing information leakage between users or VMs. This OS-level mitigation can be implemented by issuing a \texttt{TILERELEASE} or similar reset instruction during task switching, which guarantees that each process starts from the coldest AMX state. In scenarios where AMX-based models are co-resident (e.g., MLaaS environments, containers, enclaves), this prevents reuse-based side channels. However, this comes at a dynamic cost: if context switches are frequent, the cold-start latency is repeatedly reintroduced.\textit{ At low switching rates (e.g., <10 switches/sec), this overhead is negligible—less than 2\% for both power and performance}. 

But as shown in  Figure~\ref{fig:overhead}, as the switch rate approaches 1000/sec, power cost climbs to 11.6\% and performance overhead reaches 30\%, closely matching the always-cold extremes. Unlike fixed-stage defenses, however, this approach maintains AMX’s power-saving behavior for workloads that are not switching often. This provides a tunable tradeoff for secure, multi-tenant systems, with low impact in realistic workloads. This OS-level strategy offers a practical, efficient mitigation for shared-core deployments, isolating AMX timing state without permanent power-on cost.

For workloads with predictable AMX usage patterns, compiler support could inject dummy AMX instructions in conditional paths to maintain constant-time behavior. This compiler-level padding can be selectively applied only to known leakage-prone structures such as MoE dispatch and early-exit classifiers. Finally, hardware vendors should consider integrating a secure runtime control plane that allows compilers or OSes to set AMX residency policy directly—e.g., ``warm mode'', ``reset-on-swap'', or ``cold-safe''—to reflect the sensitivity and latency demands of the running code. One option is to modify the AMX microcode so that every \texttt{TMUL} instruction—regardless of prior usage—executes at a fixed latency. Alternatively, AMX’s power-gating transitions could be smoothed or disabled to keep it semi-active without a full shutdown.

Future work should refine Intel AMX power management to balance security, power, and performance, considering \scheme{} attacks.

\vspace{-1em}
\section{Conclusion}\label{sec:conclusion}
We present a security analysis of Intel AMX and reveal a novel timing vulnerability 
\scheme, which exploits reuse-distance-dependent latency caused by power gating 
to leak information across OS, VM, and enclave boundaries with high signal strength and minimal attacker control.

In the ML domain, developers of sensitive models (e.g., private or MLaaS deployments) must now consider timing leaks related to power optimization,
potentially requiring defenses like the proposed defense or model logic redesign. For instance, MoE or early-exit networks may need to be avoided or confined to low-risk contexts. 
The discovery of gadgets in widely used frameworks means library maintainers may need to patch the relevant code. 
OS and hardware manufacturers may need to get updated and issue guidance (e.g., resetting AMX during context switch, inserting dummy AMX ops, or disabling AMX within enclaves) and consider more secure designs for Intel AMX power optimization to mitigate this risk.


\begin{acks}
The authors thank anonymous reviewers for their helpful comments and feedback. This work was supported by Semiconductor Research Corporation (SRC) contract \#2025-HW-3306 and Intel Labs. 
\end{acks}

\bibliographystyle{ACM-Reference-Format}
\bibliography{refs}

\end{document}